# A non-hybrid method for the PDF equations of turbulent flows on unstructured grids

J. Bakosi *, P. Franzese, and Z. Boybeyi

*College of Science, George Mason University, Fairfax, VA, 22030, USA*

**Abstract**

In probability density function (PDF) methods of turbulent flows, the joint PDF of several flow variables is computed by numerically integrating a system of stochastic differential equations for Lagrangian particles. A set of parallel algorithms is proposed to provide an efficient solution of the PDF transport equation modeling the joint PDF of turbulent velocity, frequency and concentration of a passive scalar in geometrically complex configurations. In the vicinity of walls the flow is resolved by an elliptic relaxation technique down to the viscous sublayer, explicitly modeling the high anisotropy and inhomogeneity of the low-Reynolds-number wall region without damping or wall-functions. An unstructured Eulerian grid is employed to extract Eulerian statistics, to solve for quantities represented at fixed locations of the domain (i.e. the mean pressure and the elliptic relaxation tensor) and to track particles. All three aspects regarding the grid make use of the finite element method (FEM) employing the simplest linear FEM shapefunctions. To model the small-scale mixing of the transported scalar, the interaction by exchange with the conditional mean (IECM) model is adopted. An adaptive algorithm that computes the velocity-conditioned scalar mean is proposed that homogenizes the statistical error over the sample space with no assumption on the shape of the underlying velocity PDF. Compared to other hybrid particle-in-cell approaches for the PDF equations, the current methodology is consistent without the need for consistency conditions. The algorithm is tested by computing the dispersion of passive scalars released from concentrated sources in two different turbulent flows: the fully developed turbulent channel flow and a street canyon (or cavity) flow. Algorithmic details on estimating conditional and unconditional statistics, particle tracking and particle-number control are presented in detail. Relevant aspects of performance and parallelism on cache-based shared memory machines are discussed.

*Key words:* probability density function method; particle-in-cell method; Langevin equation; Monte-Carlo method; finite element method; particle tracking; turbulent flow; unstructured grids; scalar dispersion

## 1. Introduction

Probability density function (PDF) methods (1; 2; 3) have been developed as an alternative approach to moment closure techniques to simulate turbulent flows with a higher level of statistical description. While traditional moment closures (such as the $k$–$\varepsilon$ method (4) or various Reynolds stress and related models (5; 6; 7; 8)) seek to directly determine the mean and variance of the underlying turbulent velocity field, the aim of PDF methods is different. Instead of computing statistical moments (e.g. mean momentum and

---

* Corresponding author. Tel.: +1-703-993-9283. Fax: +1-703-993-9280.
   *Email address:* `jbakosi@gmu.edu` (J. Bakosi).

Reynolds stresses) explicitly, the full probability density function is sought, which in turn can provide higher order moments if necessary. Shifting the problem to a higher level is beneficial in a number of ways. For example, one-point statistics (such as advection and chemical reaction) appear in mathematically exact form in the PDF transport equations, thus closure assumptions for these terms are not needed. The problem is most severe in chemically reacting turbulent flows, where previous attempts to provide moment closures for the usually highly nonlinear chemical source terms resulted in errors of several orders of magnitude (9). In PDF methods, the closure problem is not eliminated, since two-point statistics (such as dissipation) still require modeling assumptions. Nevertheless, since fundamental physical processes are treated exactly, a more accurate representation can be achieved. A higher level statistical description also provides more information that can be used in the construction of closure models.

The development of PDF methods has mostly been centered on chemically reactive turbulent flows on simple geometries (10; 11), although applications to more complex configurations (12; 13) as well as to atmospheric flows (14; 15) have also appeared. A large variety of compressible and incompressible laminar flows bounded by bodies of complex geometries have been successfully computed using unstructured grids (16). The flexibility of these gridding techniques has also been exploited recently in mesoscale atmospheric modeling (17). Significant advances in automatic unstructured grid generation (18), sophisticated data structures and algorithms, automatic grid refinement and coarsening techniques (19) in recent years have made unstructured grids a common and convenient choice of spatial discretization in computational physics. The success of unstructured grids seems to warrant exploiting their advantages in conjunction with PDF modeling. For reasons to be elaborated on later, in PDF methods the usual choice of representation is the Lagrangian framework with a numerical method employing a large number of Lagrangian particles. A natural way to combine the advantages of existing traditional Eulerian computational fluid dynamics (CFD) codes with PDF methods, therefore, is to develop hybrid methods.

Using structured grids, a hybrid finite-volume (FV)/particle method has been developed by Muradoglu *et al.* (20) and Jenny *et al.* (21), wherein the mean velocity and pressure fields are supplied by the FV code to the particle code, which in turn obtains the Reynolds stress, scalar fluxes and reaction terms. Different types of hybrid algorithms are possible depending on which quantities are computed in the Eulerian and Lagrangian frameworks. For a list of approaches see (20). Another line of research has been centered on the combination of large eddy simulation (LES) with PDF methods (22; 23). This approach is based on the definition of the filtered density function (FDF) (9) which is used to provide closure at the residual scale to the filtered LES equations. Depending on the flow variables included in the joint FDF, different variants of the method have been proposed providing a probabilistic treatment at the residual scale for species compositions (24), velocity (25) and velocity and scalars (26). A common feature of these methods is that certain consistency conditions have to be met, since some fields are computed in both the Eulerian and Lagrangian frameworks. Further advances on consistency conditions and correction algorithms for hybrid FV/particle codes have been reported by Muradoglu *et al.* (27) and Zhang & Haworth (28), who also extend the hybrid formulation to unstructured grids. Following that line, a hybrid algorithm for unstructured multiblock grids has recently been proposed by Rembold & Jenny (29). Beside enforcing the consistency of redundantly computed fields, hybrid methods also have to be designed to ensure consistency at the level of the turbulence closure between the two frameworks. For example, the simplified Langevin model (SLM) (30) is equivalent to Rotta's model at the Reynolds stress level (31). Thus the use of a $k$–$\varepsilon$ model in the Eulerian framework and of a SLM PDF model in the Lagrangian framework cannot be consistent (20). In this paper, a different approach is taken by representing all turbulent fields by Lagrangian particles and employing the grid (a) to compute only inherently Eulerian quantities (that are only represented in the Eulerian sense), (b) to extract Eulerian statistics and (c) to locate particles throughout the domain. Because the resulting method is not a hybrid one, none of the fields are computed redundantly and the computation can remain fully consistent without the need of correction algorithms. We employ the finite element method (FEM) in all three aspects mentioned above in conjunction with Eulerian grids. The combined application of the FEM and the decoupling of the Eulerian and Lagrangian fields also have important consequences regarding particle boundary conditions as compared to the "flux-view" of FV methods.

In those turbulent flows where an accurate description of the velocity field is required in the vicinity of walls, adequate representation of the near-wall low-Reynolds number effects is essential. High-Reynolds-number turbulence closures, therefore, have to be adjusted close to walls. Possible modifications involve damping functions (32; 33; 34; 35) or wall-functions (36; 37; 38; 39). For those flows, where accurate higher order statistics are also required at the wall, capturing the near-wall inhomogeneity and anisotropy of the Reynolds stress tensor is crucial. Following Durbin's elliptic relaxation technique (40), Dreeben and Pope (41) extended the PDF method to include wall-treatment. In their model, only the no-slip and impermeability conditions are imposed on particles close to walls and an elliptic equation for a tensorial quantity brings out the non-local effect of the wall on the Reynolds stresses. Wall-function treatment has also been developed for the PDF framework (42), providing the option of the usual trade-off between computational expense and resolution at walls. For our current purposes, we adopt the elliptic relaxation technique and resolve the flow all the way to the wall.

Beside in chemically reactive turbulent flows, the transport and dispersion of scalars (e.g. species concentration or pollution) is a central issue in computational atmospheric physics, as well. Reviews on the subject have been compiled by Warhaft (43) and Karnik & Tavoularis (44). In addition to the velocity field, we include in our formulation the capability to model the concentration of a passive scalar released from a concentrated source, employing the interaction by exchange with the conditional mean (IECM) model to incorporate the effects of small-scale mixing on the scalar. For the computation of the velocity-conditioned scalar mean required in the IECM model, we propose an adaptive algorithm that makes no assumption on the shape of the underlying velocity PDF and which, using a dynamic procedure, automatically homogenizes the statistical error over the sample space. We extend our description of the algorithm to shared memory parallelism and highlight relevant aspects of serial and parallel efficiency.

The purpose of this research is to continue to widen the applicability of PDF methods in practical applications, especially to more realistic flow geometries by employing unstructured grids. The current work is a step in that direction, where we combine several models and develop a set of parallel algorithms to compute the joint PDF of the turbulent velocity, characteristic frequency and scalar concentration in complex domains. Complementary to hybrid FV/particle methods, we provide a different methodology to exploit the advantages of unstructured Eulerian meshes in conjunction with Lagrangian PDF methods. Two simple flows, a fully developed turbulent channel flow and a street canyon (or cavity) flow, are used to test several aspects of the algorithms. Both of these cases are two-dimensional; however, the methodology is general enough so that the extension from 2d triangles to 3d tetrahedra should be straightforward.

The paper is organized as follows. In Section 2 the governing equations are described. Section 3 presents details of the solution algorithm with the underlying numerical methods. The effects of several algorithmic characteristics on selected one-point statistics are explored on the two testcases in Section 4: scalar dispersion from concentrated sources is examined in a fully developed turbulent channel flow in Section 4.1 and in a turbulent street canyon in Section 4.2. Finally, some conclusions are drawn and future directions are outlined in Section 5.

## 2. Governing equations

The governing system of equations for a passive scalar released in a viscous, Newtonian, incompressible fluid is written in the Eulerian framework as

$$\frac{\partial U_i}{\partial x_i} = 0, \tag{1}$$

$$\frac{\partial U_i}{\partial t} + U_j \frac{\partial U_i}{\partial x_j} + \frac{1}{\rho}\frac{\partial P}{\partial x_i} = \nu \nabla^2 U_i, \tag{2}$$

$$\frac{\partial \phi}{\partial t} + U_i \frac{\partial \phi}{\partial x_i} = \Gamma \nabla^2 \phi, \tag{3}$$

where $U_i$, $P$, $\rho$, $\nu$, $\phi$ and $\Gamma$ are the Eulerian velocity, pressure, constant density, kinematic viscosity, scalar concentration and scalar diffusivity, respectively. Based on these equations an exact transport equation can be derived for the one-point, one-time Eulerian joint PDF of velocity and concentration $f(\boldsymbol{V}, \psi; \boldsymbol{x}, t)$ (2; 45),

$$\frac{\partial f}{\partial t} + V_i \frac{\partial f}{\partial x_i} = \nu \frac{\partial^2 f}{\partial x_i \partial x_i} + \frac{1}{\rho}\frac{\partial \langle P \rangle}{\partial x_i}\frac{\partial f}{\partial V_i} - \frac{\partial^2}{\partial V_i \partial V_j}\left[f\left\langle \nu \frac{\partial U_i}{\partial x_k}\frac{\partial U_j}{\partial x_k}\bigg| U = V, \phi = \psi \right\rangle\right]$$
$$+ \frac{\partial}{\partial V_i}\left[f\left\langle \frac{1}{\rho}\frac{\partial p}{\partial x_i}\bigg| U = V, \phi = \psi \right\rangle\right] - \frac{\partial}{\partial \psi}\left[f\langle \Gamma \nabla^2 \phi | U = V, \phi = \psi \rangle\right], \quad (4)$$

where $V$ and $\psi$ denote the sample space variables of the stochastic velocity $U(x,t)$ and concentration $\phi(x,t)$ fields, respectively and the pressure $P$ is decomposed into its mean $\langle P \rangle$ and fluctuation part $p$. A remarkable feature of Eq. (4) is that the effects of convection and viscous diffusion (processes of critical importance in wall-bounded turbulent flows) are in closed form, thus require no modeling assumptions. The last three terms, however, are unclosed. These are respectively, the effects of dissipation of turbulent kinetic energy, pressure redistribution and the small-scale mixing of the transported scalar due to molecular diffusion. The joint PDF $f(V, \psi; x, t)$ contains all one-point statistics of the velocity and scalar fields. The price to pay for the increased level of description (compared to traditional moment closures) is that in a general three-dimensional turbulent flow $f(V, \psi; x, t)$ is a function of 8 independent variables. This effectively rules out the application of traditional techniques like the finite difference, finite volume or finite element methods for a numerical solution. While in principle this high-dimensional space could be discretized and (after appropriate modeling of the unclosed terms) Eq. (4) could be solved with the above methods, the preferred choice in the PDF framework is to use a Lagrangian Monte-Carlo formulation. As opposed to the other techniques mentioned, the computational requirements increase only linearly with increasing problem dimension with a Monte-Carlo method. Another advantage of employing a Lagrangian-particle based simulation is that the governing equations may take a significantly simpler form than Eq. (4).

In a Lagrangian formulation, it is assumed that the motion of fluid particles along their trajectory is well represented by a diffusion process, namely a continuous-time Markov process with continuous sample paths (46). Such a process was originally proposed by Langevin (47) as a stochastic model of a microscopic particle undergoing Brownian motion. Pope (45) shows that Langevin's equation provides a good model for the velocity of a fluid particle in turbulence. It is important to appreciate that the instantaneous particle velocities modeled by a Langevin equation do not represent individual physical fluid particle velocities. Rather, their combined effect (i.e. their statistics) can model statistics of a turbulent flow. Therefore, the numerical particles can be thought of as an ensemble representation of turbulence, each particle embodying one realization of the flow at a given point in space and time. At a fundamental level, an interesting consequence of this view is that this definition does not require an external (spatial or temporal) filter explicitly, as the classical Reynolds averaging rules and large eddy simulation (LES)-filtering do. For example, in unsteady homogeneous or steady inhomogeneous high-Reynolds-number flows, the natural Reynolds-average to define is the spatial and temporal average, respectively. In unsteady *and* inhomogeneous flows however, one is restricted to employ temporal and/or spatial filters leading to the approaches of unsteady Reynolds-averaged Navier-Stokes (URANS) and LES methods, respectively. This leads to formulations (and results) that depend on the artificially introduced (and flow dependent) filter width, which is clearly not desirable (48). In the PDF framework the statistics are defined based on a probability density function. In the current case, for example, the mean velocity and Reynolds stress tensor are obtained from the joint PDF $f$ as

$$\langle U_i \rangle(x,t) \equiv \int_{-\infty}^{\infty} \int_{0}^{\infty} V_i f(V, \psi; x, t) \mathrm{d}\psi \mathrm{d}V, \quad (5)$$

$$\langle u_i u_j \rangle(x,t) \equiv \int_{-\infty}^{\infty} \int_{0}^{\infty} (V_i - \langle U_i \rangle)(V_j - \langle U_j \rangle) f(V, \psi; x, t) \mathrm{d}\psi \mathrm{d}V, \quad (6)$$

where the velocity fluctuation is defined as $u_i = V_i - \langle U_i \rangle$. These quantities are well-defined mathematically (46; 45), independently of the underlying physics, the state of the flow (i.e. homogeneous or inhomogeneous, steady or unsteady), the numerical method and the spatial and temporal discretization. Therefore the promise of a probabilistic view of turbulence (as in PDF methods) at the fundamental level is a more rigorous statistical treatment.

An equivalent model to the Eulerian momentum equation (2) in the Lagrangian framework is a system of governing equations for particle position $\mathcal{X}_i$ and velocity $\mathcal{U}_i$ increments (49)

$$d\mathcal{X}_i = \mathcal{U}_i dt + (2\nu)^{1/2} dW_i, \tag{7}$$

$$d\mathcal{U}_i(t) = -\frac{1}{\rho}\frac{\partial P}{\partial x_i}dt + 2\nu\frac{\partial^2 U_i}{\partial x_j \partial x_j}dt + (2\nu)^{1/2}\frac{\partial U_i}{\partial x_j}dW_j, \tag{8}$$

where the isotropic Wiener process $dW_i$ (50) is identical in both equations (numerically, the same exact series of Gaussian random numbers with zero mean and variance $dt$) and it is understood that the Eulerian fields on the right hand side are evaluated at the particle locations $\mathcal{X}_i$. Since Eq. (8) is a diffusion-type stochastic differential equation with a Gaussian white noise (i.e. a Wiener process), it is equivalent to the Fokker-Planck equation that governs the evolution of the probability distribution of the same process (46). Eqs. (7) and (8) represent the viscous effects exactly in the Lagrangian framework. Particles governed by these equations are both advected and diffused in physical space. After Reynolds decomposition is applied to the velocity and pressure, Eq. (8) results in

$$\begin{aligned} d\mathcal{U}_i(t) = &-\frac{1}{\rho}\frac{\partial \langle P \rangle}{\partial x_i}dt + 2\nu\frac{\partial^2 \langle U_i \rangle}{\partial x_j \partial x_j}dt + (2\nu)^{1/2}\frac{\partial \langle U_i \rangle}{\partial x_j}dW_j \\ &-\frac{1}{\rho}\frac{\partial p}{\partial x_i}dt + 2\nu\frac{\partial^2 u_i}{\partial x_j \partial x_j}dt + (2\nu)^{1/2}\frac{\partial u_i}{\partial x_j}dW_j, \end{aligned} \tag{9}$$

where the last three terms are unclosed. To model these terms, we adopt the generalized Langevin model (GLM) of Haworth & Pope (30)

$$\begin{aligned} d\mathcal{U}_i(t) = &-\frac{1}{\rho}\frac{\partial \langle P \rangle}{\partial x_i}dt + 2\nu\frac{\partial^2 \langle U_i \rangle}{\partial x_j \partial x_j}dt + (2\nu)^{1/2}\frac{\partial \langle U_i \rangle}{\partial x_j}dW_j \\ &+ G_{ij}\left(\mathcal{U}_j - \langle U_j \rangle\right)dt + (C_0 \varepsilon)^{1/2} dW'_i, \end{aligned} \tag{10}$$

where $G_{ij}$ is a second-order tensor function of velocity statistics, $C_0$ is a positive constant, $\varepsilon$ denotes the rate of dissipation of turbulent kinetic energy and $dW'_i$ is another Wiener process. Because of the correspondence between stochastic Lagrangian models and Reynolds stress closures (31), different second order models can be realized with the Langevin equation (10), depending on how $G_{ij}$ is specified. An advantage of this family of models is that the GLM equation (10) ensures realizability as a valid Reynolds stress closure, provided that $C_0$ is non-negative and that $C_0$ and $G_{ij}$ are bounded (45). Compared to Reynolds stress closures, the terms in $G_{ij}$ and $C_0$ represent pressure redistribution and anisotropic dissipation of turbulent kinetic energy. Far from walls, these physical processes can be adequately modeled by appropriate local (algebraic) functions of the velocity statistics, however, such local representation is in contradiction with the large structures interacting with the wall and the viscous wall region (51). The traditionally employed damping or wall-functions, therefore, are of limited validity in an approach aiming at a higher-level statistical description. To address these issues, Durbin (40) proposed a technique, which incorporates the wall-effects on the Reynolds stress tensor in a more natural fashion. In his approach, an elliptic equation is employed to capture the non-locality of the pressure redistribution at the wall, based on the analogy with the Poisson equation which governs the pressure in incompressible flows. The methodology also provides more freedom on controlling the individual components of the Reynolds stress tensor at the wall, such as the suppression of only the wall-normal component representing wall-blocking. Dreeben & Pope (41) incorporated Durbin's elliptic relaxation technique into the PDF method, by specifying $G_{ij}$ and $C_0$ through the tensor $\wp_{ij}$ as

$$G_{ij} = \frac{\wp_{ij} - \frac{\varepsilon}{2}\delta_{ij}}{k} \quad \text{and} \quad C_0 = \frac{-2\wp_{ij}\langle u_i u_j \rangle}{3k\varepsilon}, \tag{11}$$

where $k = \frac{1}{2}\langle u_i u_i \rangle$ denotes the turbulent kinetic energy. The non-local quantity $\wp_{ij}$ is specified with the following elliptic relaxation equation

$$\wp_{ij} - L^2 \nabla^2 \wp_{ij} = \frac{1-C_1}{2} k \langle \omega \rangle \delta_{ij} + k H_{ijkl} \frac{\partial \langle U_k \rangle}{\partial x_l}, \tag{12}$$

where the fourth-order tensor $H_{ijkl}$ is given by

$$H_{ijkl} = (C_2 A_v + \tfrac{1}{3}\gamma_5)\delta_{ik}\delta_{jl} - \tfrac{1}{3}\gamma_5 \delta_{il}\delta_{jk} + \gamma_5 b_{ik}\delta_{jl} - \gamma_5 b_{il}\delta_{jk}, \tag{13}$$

$$A_v = \min\left[1.0, C_v \frac{\det \langle u_i u_j \rangle}{\left(\tfrac{2}{3}k\right)^3}\right], \tag{14}$$

and

$$b_{ij} = \frac{\langle u_i u_j \rangle}{\langle u_k u_k \rangle} - \frac{1}{3}\delta_{ij} \tag{15}$$

is the Reynolds stress anisotropy, $\langle \omega \rangle$ denotes the mean characteristic turbulent frequency and $C_1, C_2, \gamma_5, C_v$ are model constants. The characteristic lengthscale $L$ is defined by the maximum of the turbulent and Kolmogorov lengthscales

$$L = C_L \max\left[C_\xi \frac{k^{3/2}}{\varepsilon}, C_\eta \left(\frac{\nu^3}{\varepsilon}\right)^{1/4}\right], \tag{16}$$

with

$$C_\xi = 1.0 + 1.3 n_i n_i, \tag{17}$$

where $n_i$ is the unit wall-normal of the closest wall-element pointing outward of the flow domain, while $C_L$ and $C_\eta$ are model constants. The right hand side of Eq. (12) can be any local model for pressure redistribution; here we follow Dreeben & Pope (41) and use the stochastic Lagrangian equivalent of a modified isotropization of production (IP) model proposed by Pope (31). It is apparent that Eq. (12) acts like a blending function between the low-Reynolds-number near-wall region and the high-Reynolds-number free turbulence. Close to the wall, the elliptic term on the left hand side brings out the non-local, highly anisotropic behavior of the Reynolds stress tensor, whereas far from the wall the significance of the elliptic term vanishes and the local model on the right hand side is recovered. A difference compared to the original PDF model is the application of the elliptic term $L^2 \nabla^2 \wp_{ij}$ as proposed originally by Durbin (40), as opposed to $L\nabla^2(L\wp_{ij})$, since no visible improvement has been found by employing the latter, numerically more expensive term.

The description of the computation of the mean-pressure gradient in Eq. (10) is deferred to Section 3.2. We complete the closure of Eq. (10) by specifying the turbulent kinetic energy dissipation rate $\varepsilon$ as (41)

$$\varepsilon = \langle \omega \rangle \left(k + \nu C_T^2 \langle \omega \rangle\right), \tag{18}$$

where $C_T$ is a model constant and the stochastic turbulent frequency $\omega$ is calculated employing the model of van Slooten *et al.* (52)

$$d\omega = -C_3 \langle \omega \rangle \left(\omega - \langle \omega \rangle\right) dt - S_\omega \langle \omega \rangle \omega dt + \left(2 C_3 C_4 \langle \omega \rangle^2 \omega\right)^{1/2} dW, \tag{19}$$

where $dW$ is a scalar valued Wiener-process and $S_\omega$ is a source/sink term for the mean turbulent frequency

$$S_\omega = C_{\omega 2} - C_{\omega 1} \frac{\mathcal{P}}{\varepsilon}, \tag{20}$$

where $\mathcal{P} = -\langle u_i u_j \rangle \partial \langle U_i \rangle / \partial x_j$ is the production of turbulent kinetic energy and $C_3, C_4, C_{\omega 1}$ and $C_{\omega 2}$ are model constants. A simplification of the original model for the turbulent frequency employed by Dreeben & Pope (41) is the elimination of the ad-hoc source term involving an additional model constant, since in our case studies we found no obvious improvements by including it. This completes the model for the joint PDF of velocity and the (now included) characteristic turbulent frequency $\omega$.

Since a passive scalar, by definition, has no effect on the turbulent velocity field, modeling the pressure redistribution and dissipation have been discussed independently from the scalar, i.e. it has been assumed that in Eq. (4) the following hold

$$\left\langle \nu \frac{\partial U_i}{\partial x_k} \frac{\partial U_j}{\partial x_k} \middle| \boldsymbol{U} = \boldsymbol{V}, \phi = \psi \right\rangle = \left\langle \nu \frac{\partial U_i}{\partial x_k} \frac{\partial U_j}{\partial x_k} \middle| \boldsymbol{U} = \boldsymbol{V} \right\rangle, \tag{21}$$

$$\left\langle \frac{1}{\rho} \frac{\partial p}{\partial x_i} \middle| \boldsymbol{U} = \boldsymbol{V}, \phi = \psi \right\rangle = \left\langle \frac{1}{\rho} \frac{\partial p}{\partial x_i} \middle| \boldsymbol{U} = \boldsymbol{V} \right\rangle. \tag{22}$$

However, the opposite, that the micromixing of the scalar can be modeled independently of $\boldsymbol{V}$, cannot be assumed in general (53). A simple mixing model is the interaction by exchange with the mean (IEM) (54; 55), which models the conditional scalar diffusion in Eq. (4) independent of the underlying velocity field i.e. assuming $\langle \Gamma \nabla^2 \phi | \boldsymbol{U} = \boldsymbol{V}, \phi = \psi \rangle = \langle \Gamma \nabla^2 \phi | \phi = \psi \rangle$ in Eq. (4). In the Lagrangian framework, the IEM model is written as

$$\mathrm{d}\psi = -\frac{1}{t_\mathrm{m}} (\psi - \langle \phi \rangle) \, \mathrm{d}t, \tag{23}$$

where $t_\mathrm{m}$ is a micromixing timescale. It has been pointed out, however, that the assumption that the scalar mixing is independent of the velocity bears no theoretical justification and is at odds with local isotropy of the scalar field (56; 53). On the other hand, the interaction by exchange with the conditional mean (IECM) model does take the velocity field into consideration by employing the velocity-conditioned mean instead of the unconditional mean as

$$\mathrm{d}\psi = -\frac{1}{t_\mathrm{m}} (\psi - \langle \phi | \boldsymbol{U} = \boldsymbol{V} \rangle) \, \mathrm{d}t. \tag{24}$$

This model represents the physical process of dissipation by relaxation of the particle concentration $\psi$ towards the conditional scalar mean with timescale $t_\mathrm{m}$. It can be shown that in the case of homogeneous turbulent mixing with no mean scalar gradient the IEM and IECM models are equivalent since the velocity and scalar fields are uncorrelated (56). In that case the micromixing timescale $t_\mathrm{m}$ is proportional to the Kolmogorov timescale $\tau = k/\varepsilon$. In the current study we focus on transported scalars released from concentrated sources in flow domains surrounded by no-slip walls, thus we expect the scalar fields to be highly inhomogeneous. Accordingly, we follow (57) and specify the micromixing timescale as a function of the location $\boldsymbol{r}$

$$t_\mathrm{m}(\boldsymbol{r}) = \min \left[ C_s \left( \frac{r_0^2}{\varepsilon} \right)^{1/3} + C_t \frac{d_{\boldsymbol{r}}}{U_c(\boldsymbol{r})}; \ \max \left( \frac{k}{\varepsilon}, C_T \sqrt{\frac{\nu}{\varepsilon}} \right) \right], \tag{25}$$

where $r_0$ denotes the radius of the source, $U_c$ is a characteristic velocity at $\boldsymbol{r}$ which we take as the absolute value of the mean velocity at the given location, $d_{\boldsymbol{r}}$ is the distance of the point $\boldsymbol{r}$ from the source, while $C_s$ and $C_t$ are model constants.

This completes the model for the joint PDF of turbulent velocity, frequency and scalar. The model is 'complete' in the sense that the equations are free from flow-dependent specifications (45). Thus, in principle, it is generally applicable to any transported passive scalar released into an incompressible, high-Reynolds-number flow from a concentrated source. To summarize, the flow is modeled by a large number of Lagrangian particles representing a finite sample of all fluid particles, which can be thought of as different realizations of the underlying stochastic fields. Consequently, employing appropriate ensemble averages, all one-point statistics can be obtained. Numerically, each particle has its position $\mathcal{X}_i$ and with its velocity $\mathcal{U}_i$ carries its characteristic frequency $\omega$ and scalar concentration $\psi$. These quantities are advanced by Eqs. (7), (10), (19) and (24), respectively.

## 3. Numerical implementation

The numerical solution algorithm is based on the time-dependent particle governing equations (7), (10), (19) and (24). An adaptive timestepping strategy to advance the system is described in Section 3.1. All Eulerian statistics required in these equations need to be estimated at the particle locations at the given instant in time. This is performed by the use of an unstructured Eulerian grid that discretizes the flow domain, which can be conveniently refined around regions where a higher resolution is necessary. The methods used to compute unconditional statistics, their derivatives and conditional statistics are described in Sections 3.3, 3.4 and 3.5, respectively. The grid is also used to solve the elliptic relaxation equation (12) and to solve for the mean pressure required in Eq. (10). The main characteristics of the solution of these two Eulerian equations together with a projection method to obtain the mean pressure are described in Section 3.2. In order to identify which particles contribute to local statistics, the particles need to be continuously followed as they travel throughout the domain. The particle tracking algorithm that is used for this purpose is described in Section 3.6. In complex configurations, where the spatial resolution can differ significantly from one region to another, an algorithm is necessary to ensure that the number of particles in every computational element is above a certain threshold, so meaningful statistics can be computed. We present an algorithm that accomplishes this task in Section 3.7. The boundary conditions at no-slip walls applied to particles, to the elliptic relaxation equation (12) and to the mean pressure are described in Section 3.8. Some aspects of parallel random number generation are described in Section 3.9. An overview of the solution procedure with the execution profile of a timestep is given in Section 3.10.

### 3.1. *Timestepping procedure*

To discretize in time the governing equations (7), (10), (19), (24) we apply explicit forward Euler timestepping. The size of the timestep is estimated in every step based on the Courant-Friedrichs-Lewy (CFL) (58) condition as

$$\Delta t = \text{CFL} \frac{\min_n \sqrt{A_n}}{\max_n ||\langle \mathbf{Z} \rangle_n||_2}, \tag{26}$$

where $A_n$ is the average element area around gridnode $n$ and $\mathbf{Z} = (V_1, V_2, V_3, \omega, \psi)$ is the combined vector of particle properties. According to Eq. (26) we find the smallest characteristic edge length (defined by the square-root of the element area) on the whole domain and divide it by the largest characteristic velocity (based on the length of the generalized mean velocity vector $\langle \mathbf{Z} \rangle$). This conservative approximation is multiplied by a CFL constant of 0.7.

### 3.2. *Solution of the Eulerian equations: mean pressure and elliptic relaxation*

In incompressible flows the pressure establishes itself immediately through the pressure-Poisson equation, which is a manifestation of the divergence constraint $\nabla \cdot \mathbf{U} = 0$ expressing mass conservation. The numerical difficulties arising from the straightforward discretization of this equation in finite difference, finite volume and finite element methods are reviewed in (16). Several different methods have been devised to deal with these issues, which stem from the fact that the mass conservation equation decouples from the momentum equation and acts on it only as a constrain, which may result in the decoupling of every second gridpoint thereby numerically destabilizing the solution. Some of these methods are: the use of different functional spaces for the velocity and pressure discretization, artificial viscosities, consistent numerical fluxes, artificial compressibility and pressure projection schemes. For our purposes we adopt the pressure projection approach.

Additionally, due to the stochastic nature of the simulation, in PDF methods the Eulerian statistics and their derivatives are subject to considerable statistical noise. Fox (59) suggests three different ways of calculating the mean pressure in PDF methods. The first approach is to extract the mean pressure field from a simultaneous consistent Reynolds stress model solved using a standard CFD solver (60). This approach solves the noise problem although it leads to a redundancy in the velocity model. The second approach attacks the noise problem by computing the so-called 'particle-pressure field' (61). This results in a stand-alone transported PDF method and the authors succesfully apply it to compute a compressible turbulent flow. The third approach is the hybrid methodology, which uses an Eulerian CFD solver to solve for the mean velocity field and a particle-based code to solve for the fluctuating velocity (20). This method is made consistent by the careful selection of turbulence models in the Eulerian and Lagrangian frameworks and the use of consistency conditions.

A different approach is proposed here. We adopt a modified version of the pressure projection scheme originally proposed by Chorin (62) in the finite difference context, which has been widely used in laminar flows. The modification compared to the original projection scheme involves solving for the difference of the pressure between two consecutive timesteps, instead of the pressure field itself. This ensures that at steady state the residuals of the pressure correction vanish (16). We adopt the scheme in the Lagrangian-Eulerian setting and combine the projection algorithm with the particle equations as follows.

The idea of pressure projection is to first predict the velocity using the current flow variables without taking the divergence constraint into consideration. In a second step, the divergence constraint is enforced by solving a pressure-Poisson equation. Finally the velocity is corrected using the new pressure field, resulting in a divergence-free velocity-field. Thus, for an explicit (forward Euler) time-integration of the particle velocity, one complete timestep ($n \to n+1$) is given by:

– *Velocity prediction:* $\mathcal{U}^n \to \mathcal{U}^*$

$$\mathcal{U}_i^* = \mathcal{U}_i^n - \frac{1}{\rho}\frac{\partial \langle P \rangle^n}{\partial x_i}\Delta t + 2\nu \frac{\partial^2 \langle U_i \rangle^n}{\partial x_j \partial x_j}\Delta t + (2\nu)^{1/2}\frac{\partial \langle U_i \rangle^n}{\partial x_j}\Delta W_j \\ + G_{ij}\left(\mathcal{U}_j^n - \langle U_j \rangle^n\right)\Delta t + (C_0\varepsilon)^{1/2}\Delta W_i'; \tag{27}$$

– *Pressure projection:* $\langle P \rangle^n \to \langle P \rangle^{n+1}$

$$\nabla \cdot \langle \bm{U} \rangle^{n+1} = 0, \tag{28}$$

$$\frac{\langle \bm{U} \rangle^{n+1} - \langle \bm{U} \rangle^*}{\Delta t} + \frac{1}{\rho}\nabla(\langle P \rangle^{n+1} - \langle P \rangle^n) = 0, \tag{29}$$

which results in

$$\frac{1}{\rho}\nabla^2(\langle P \rangle^{n+1} - \langle P \rangle^n) = \frac{\nabla \cdot \langle \bm{U} \rangle^*}{\Delta t}; \tag{30}$$

– *Mean velocity correction:* $\langle \bm{U} \rangle^* \to \langle \bm{U} \rangle^{n+1}$

$$\langle \bm{U} \rangle^{n+1} = \langle \bm{U} \rangle^* - \frac{1}{\rho}\Delta t \nabla(\langle P \rangle^{n+1} - \langle P \rangle^n). \tag{31}$$

Since the velocity field is fully represented by particles, the velocity prediction (27) and correction (31) steps are applied to particles. The above procedure ensures that the Poisson equation for the mean pressure is satisfied at all times, thus the joint PDF representing an incompressible flow satisfies realizability, normalization and consistency conditions (2) in every timestep. To stabilize the computation of the mean pressure a small artificial diffusion term is added to the divergence constraint in Eq. (28)

$$\nabla \cdot \langle \bm{U} \rangle^{n+1} = C_p \frac{1}{\rho}\nabla^2 \langle P \rangle^n, \tag{32}$$

where $C_p$ is a small constant, e.g. $C_p = 10^{-3}$, which results in the stabilized version of the pressure projection step

$$\frac{1}{\rho}\nabla^2(\langle P \rangle^{n+1} - \langle P \rangle^n) = \frac{1}{\Delta t}\left(\nabla \cdot \langle \bm{U} \rangle^* - C_p\frac{1}{\rho}\nabla^2 \langle P \rangle^n\right). \tag{33}$$

Both the elliptic relaxation (12) and pressure projection (33) equations are solved with the finite element method using linear shapefunctions on a grid consisting of triangles (16). The FEM coefficient matrices are stored in block compressed sparse row format (63). The resulting linear systems are solved by the method of conjugate gradients combined with a Jacobi preconditioner. While the elliptic equation (12) for the tensor $\wp_{ij}$ may appear prohibitively expensive for larger meshes, the equation is well-conditioned and the iterative solution converges in a few iterations starting from an initial condition using the solution in the previous timestep.

### 3.3. *Estimation of Eulerian statistics*

During the numerical solution of the governing equations, Eulerian statistics need to be estimated at different locations of the domain. Since the joint PDF contains information on all one-point statistics of the velocity, frequency and scalar concentration fields, these are readily available through appropriate averages of particle properties. For example, the mean velocity at a specific location in space and time is obtained as the integral over all sample space of the joint PDF $\tilde{f}(\mathbf{Z})$

$$\langle U_i \rangle \equiv \int V_i \tilde{f}(\mathbf{Z}; \mathbf{x}, t) \mathrm{d}\mathbf{Z}, \tag{34}$$

where $\mathbf{Z}$ denotes the vector of all sample space variables $\mathbf{Z} = (V_1, V_2, V_3, \omega, \psi)$. For brevity we omit (but assume) the space and time dependence of the statistics. In traditional particle-codes the estimation of statistics is usually performed by kernel estimation using weight-functions (45). In particle-in-cell methods (64) an Eulerian mesh covers the computational domain and means are computed in each element or grid-point. The latter approach is followed here and Eq. (34) is computed by an ensemble average over all particle velocities in the vicinity of $\mathbf{x}$

$$\langle U_i \rangle \cong \frac{1}{N} \sum_{p=1}^{N} \mathcal{U}_i^p, \tag{35}$$

where $N$ is the number of particles participating in the local mean at $\mathbf{x}$ and $\mathcal{U}_i^p$ is the velocity vector of particle $p$. In the first pass an element-based mean is computed considering the particles in a given element. In the second pass, these element-based means are transferred to nodes of the grid by calculating the average of the elements surrounding the nodes. Wherever Eulerian statistics are needed at particle locations, like in Eq. (10), the average of the nodal values are used for all particles residing in a given element. These node-based statistics are also used in the elliptic relaxation (12) and pressure projection (33) equations. An advantage of this two-pass procedure is that a natural smoothing is inherent in transferring statistics from elements to nodes. Using only nodal statistics to update particles also makes the method more robust, since it provides an efficient guard against the unwanted occurrence of empty elements, i.e. elements without any particles. The problem of high statistical error caused by an empty element is mitigated by the other elements surrounding the given node. Linked lists (16) provide an efficient access of unstructured-grid-based data from memory (e.g. elements surrounding points, points surrounding points, etc.). Once first-order statistics, like the mean velocity, are computed, higher order statistics are calculated by the same procedure. As an example, the Reynolds stress tensor is obtained by

$$\langle u_i u_j \rangle \equiv \int (V_i - \langle U_i \rangle)(V_j - \langle U_j \rangle) \tilde{f}(\mathbf{Z}) \mathrm{d}\mathbf{Z} \cong \frac{1}{N} \sum_{p=1}^{N} \left( \mathcal{U}_i^p - \langle U_i \rangle \right) \left( \mathcal{U}_j^p - \langle U_j \rangle \right). \tag{36}$$

### 3.4. *Derivatives of Eulerian statistics*

From finite element approximation theory, an unkown function $q(\mathbf{x})$ given in nodes can be approximated over an element as

$$q(\mathbf{x}) = \sum_{j=1}^{n} N^j(\mathbf{x}) \hat{q}_j, \tag{37}$$

where $n$ is the number of nodes of the element, $\hat{q}_j$ is the value of the function $q$ in node $j$ and $N^j$ are finite element shapefunctions. For speed and simplicity, we use only a single type of element (triangle) with linear shapefunctions, which are written in the local $(\xi, \eta)$ coordinate system of the element as (see also Figure 1)

$$\begin{aligned} N^{\mathrm{A}} &= 1 - \xi - \eta, \\ N^{\mathrm{B}} &= \xi, \\ N^{\mathrm{C}} &= \eta. \end{aligned} \tag{38}$$

Employing the approximation in Eq. (37), the spatial gradient of the expectation of any function $Q(\boldsymbol{Z};\boldsymbol{x},t)$ can be computed over an element as

$$\frac{\partial Q}{\partial x_i} = \sum_{j=1}^{n} \frac{\partial N^j}{\partial x_i} \hat{Q}_j, \tag{39}$$

where $\hat{Q}_j$ denotes the nodal value of $Q$ at gridpoint $j$ of the element. The derivatives of the linear shape-functions in Eq. (38) in the global $(x,y)$ coordinate system can be derived analytically (16)

$$\frac{\partial}{\partial x}\begin{bmatrix} N^{\text{A}} \\ N^{\text{B}} \\ N^{\text{C}} \end{bmatrix} = \frac{1}{2A_e}\begin{bmatrix} -y_{\text{CA}} + y_{\text{BA}} \\ y_{\text{CA}} \\ -y_{\text{BA}} \end{bmatrix}, \qquad \frac{\partial}{\partial y}\begin{bmatrix} N^{\text{A}} \\ N^{\text{B}} \\ N^{\text{C}} \end{bmatrix} = \frac{1}{2A_e}\begin{bmatrix} x_{\text{CA}} - x_{\text{BA}} \\ -x_{\text{CA}} \\ x_{\text{BA}} \end{bmatrix}, \tag{40}$$

where $A_e$ is the area of element $e$. The derivatives are constant functions and are based only on the location of the gridpoints (see also Figure 1), e.g. $y_{\text{CA}} = y_{\text{C}} - y_{\text{A}}$. If the grid does not change during computation, these derivatives can be precomputed and stored in advance of timestepping.

Second derivatives are obtained using a two-pass procedure. In the first pass the first derivatives (which are constant over the element) are computed using Eq. (39) and then transferred to nodes by computing the averages of the elements surrounding nodes. The same procedure is applied to the derivatives in gridpoints in the second pass to obtain second derivatives.

3.5. *Estimation of the velocity-conditioned scalar mean*

Eq. (24) requires the estimation of the scalar mean conditioned on the velocity field $\langle \phi | \boldsymbol{U} = \boldsymbol{V} \rangle$ or $\langle \phi | \boldsymbol{V} \rangle$ for short. In the current case, this is defined as

$$\langle \phi | \boldsymbol{V} \rangle \equiv \int \psi \tilde{f}(\omega, \psi | \boldsymbol{V}) \mathrm{d}\omega \mathrm{d}\psi, \tag{41}$$

where the conditional PDF is usually expressed through Bayes' rule using the full PDF $\tilde{f}(\boldsymbol{V}, \omega, \psi)$ and the marginal PDF of the velocity $\tilde{f}(\boldsymbol{V})$ as

$$\tilde{f}(\omega, \psi | \boldsymbol{V}) \equiv \frac{\tilde{f}(\boldsymbol{V}, \omega, \psi)}{\tilde{f}(\boldsymbol{V})}. \tag{42}$$

Mathematically, the conditional mean $\langle \phi | \boldsymbol{V} \rangle$ defines a mean value for each combination of its conditional variables, i.e. in a three-dimensional flow, in every spatial and temporal location $\langle \phi | \boldsymbol{V} \rangle$ is a function that associates a scalar value to a vector, $\langle \phi | \boldsymbol{V} \rangle : \Re^3 \to \Re$. In practice, this means that the velocity-sample space needs to be discretized (divided into bins) and different mean values have to be computed for each bin using the particles whose velocities fall into the bin. In order to keep the statistical error small this procedure would require a large number of particles in every element. To overcome this difficulty, Fox (56) proposed a method in which the three-dimensional velocity space is projected onto a one-dimensional subspace where the discretization is carried out. This substantially reduces the need for an extensive number of particles. This projection method is exact in homogeneous turbulent shear flows, where the joint velocity PDF is Gaussian. Nevertheless, in more complex situations it can still be incorporated as a modeling assumption.

A more general way of computing the conditional mean is to use three-dimensional binning of the velociy sample space $\boldsymbol{V}$. In order to homogenize the statistical error over the sample space, the endpoints of the conditioning bins in each direction can be determined so that the distribution of the number of particles falling into the bins is as homogeneous as possible. For a Gaussian velocity PDF this can be accomplished by using statistical tables to define the endpoints (56). If the underlying velocity PDF is not known, however, another strategy is required. Note that there is absolutely no restriction on the distribution of the conditioning intervals. In other words they need not be equidistant, need not be the same (or even the same number) in every dimension and can also vary from element to element. Only some sort of clustering of the particles is needed, i.e. grouping them into subgroups of particles with similar velocities. A simple algorithm that accomplishes this task is as follows. Without loss of generality, we assume that a sample-space binning of $(2 \times 2 \times 2)$ is desired. In a first step all particles residing in the given element are sorted according to their $\mathcal{U}$ velocity component. Then the first and the second halves of the group are separately sorted according to their $\mathcal{V}$ component. After further dividing both halves into halves again, each quarter is sorted according to the $\mathcal{W}$ component. Finally, halving the quarters once again we compute scalar means for each of these 8 subgroups. Naturally, the binning can be any other structure with higher (even unequal) number of bins if that is desirable, e.g. $(5 \times 5 \times 5)$ or $(4 \times 12 \times 5)$. This procedure defines the bins dynamically based on the criterion that the bin-distribution of the number of particles be as homogeneous as possible. By doing that, it homogenizes the statistical error over the sample space and also ensures that every bin will contain particles. This simple procedure is completely general, independent of the shape and extent of the velocity PDF and dynamically adjusts the bin-distribution to the underlying PDF in every element. It is also robust, since if the number of particles in an element happens to be very low compared to the desired binning, e.g. we only have 5 particles for the 125 bins of a $(5 \times 5 \times 5)$ binning structure, the above sorting & dividing procedure can be stopped at any stage and the subgroups defined up to that stage can already be used to estimate the conditioned means. In other words, if in the above example we require that at least 2 particles should remain in every subgroup we simply stop after the first sort and only use two groups. An algorithm that accomplishes the conditioning step after the particles have been sorted into subgroups is detailed in Section 6.

### 3.6. Particle tracking

Particles have to be tracked continuously as they travel throughout the grid in order to identify which element they contribute to when local statistics are computed. A variety of algorithms with different characteristics have been developed to accomplish this task (64). Since we use explicit timestepping, the particles will not jump over many elements in a timestep, thus the fastest way to track particles is some sort of known-vicinity algorithm (65). The two-dimensional particle tracking employed here is as follows. If a particle is not in its old element (where it was in the last timestep), it is searched in the *next best element* of the surrounding elements. The knowledge of the next best element is a feature of the basic interpolation algorithm that is used to decide whether the particle resides in a given element. The interpolation algorithm is based on FEM shapefunctions, which are usually employed for approximating unknowns over elements (as it is used in Section 3.2 to discretize the Eulerian equations and in Section 3.4 to approximate functions and their derivatives) and correspond to a linear mapping between the global and local coordinates of the element, see also Figure 1. We use these shapefunctions here for interpolation in two dimensions, but this procedure can also be used in a three-dimensional case with tetrahedra (65). In the current two-dimensional case, evaluating two shapefunctions is sufficient to decide whether the particle is inside of the element. The decision is made by the following condition (see also Figure 1)

$$
\texttt{if } \left\{ \left(N^{\mathrm{A}} > 0\right) \quad \texttt{and} \quad \left(N^{\mathrm{C}} > 0\right) \quad \texttt{and} \quad \left(N^{\mathrm{A}} + N^{\mathrm{C}}\right) < A_e \right\} \quad (43)
$$
$$\quad \text{inside}$$
$$\texttt{else}$$
$$\quad \text{outside}$$

where $A_e$ is the total area of the element, while $N^{\mathrm{A}}$ and $N^{\mathrm{C}}$ are the signed half-lengths of the cross-products

$$N^{\mathrm{A}} = \frac{1}{2}\big|(\boldsymbol{r}_{\mathrm{C}} - \boldsymbol{r}_{\mathrm{B}}) \times (\boldsymbol{r}_{\mathrm{P}} - \boldsymbol{r}_{\mathrm{B}})\big|, \qquad (44)$$
$$N^{\mathrm{C}} = \frac{1}{2}\big|(\boldsymbol{r}_{\mathrm{P}} - \boldsymbol{r}_{\mathrm{B}}) \times (\boldsymbol{r}_{\mathrm{A}} - \boldsymbol{r}_{\mathrm{B}})\big|. \qquad (45)$$

Note that these are also the area coordinates of the triangle corresponding to the nodes $A$ and $C$ and also the values of the finite element shapefunctions corresponding to the three nodes, Eq. (38), evaluated at the particle location **P**. A convenient feature of this procedure is that once the values $N^A$, $N^C$ and $N^B = A_e - N^A - N^C$ are evaluated, in case the particle is not found in the element, they also give us a hint about the direction of the particle location that is outside of the element. If condition (43) is not satisfied, at least one of $N^A$, $N^B$ and $N^C$ is negative. The next best element is in the direction corresponding to the lowest of the three values. Combining this with a data structure (e.g. a linked list (16)) that stores the element indices surrounding elements, we can easily and efficiently identify which element is most likely to contain the particle or at least which direction to search next. Most of the time, the particles do not jump out of their host elements, but if they do, this procedure finds them in usually 2-3 steps.

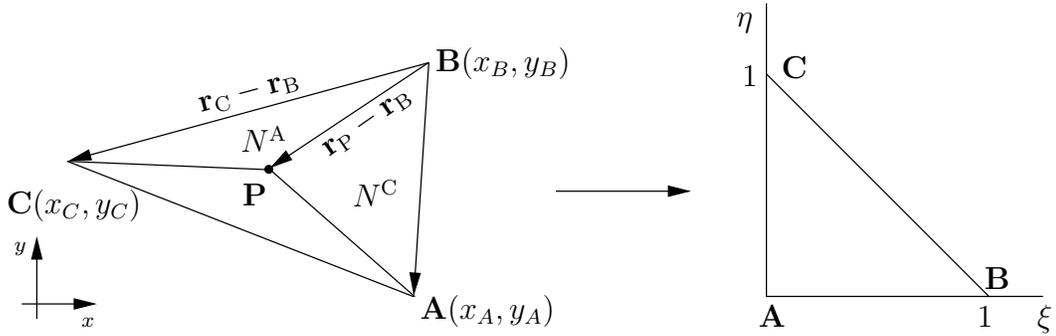

Fig. 1. The decision whether a particle resides in a triangular element is made based on computing cross-products of element-edge vectors and vectors of vertex-particle coordinates. E.g. $N^A$ is half of the signed area of the parallelogram spanned by vectors $(\boldsymbol{r}_C - \boldsymbol{r}_B)$ and $(\boldsymbol{r}_P - \boldsymbol{r}_B)$. Also shown is the local coordinate system $(\xi, \eta)$ of the triangle after a linear mapping with the finite element shapefunctions in Eqs. (38).

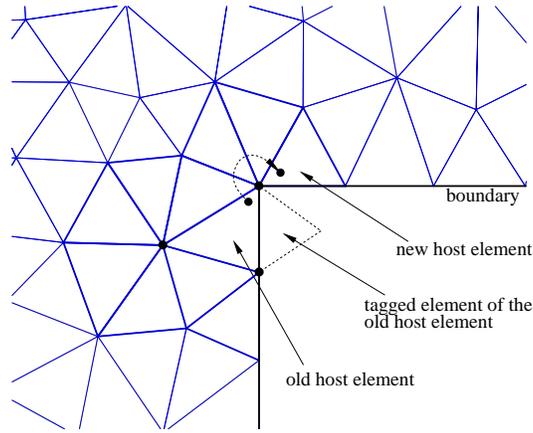

Fig. 2. A particle jumping over a concave corner on the boundary and the next best guess based on its old host element would be through the boundary, outside of the domain. A fall-back procedure finds the new host element of the particle by searching the elements surrounding the nodes (displayed with thicker edges) of its old host element.

The above neighbor-to-neighbor algorithm performs very well in the domain, but it may fail to jump over concave boundaries, resulting in a dead-lock (65). In order to remedy this problem the following strategy is employed. An element on the boundary has two surrounding elements at most and the ones that would be outside of the domain are tagged in the data structure that stores the three element indices surrounding elements, see also Figure 2. If this tagged element is returned as the next best guess, the particle is on the other side of a concave section (or a corner) of the boundary. Since even in this case the particle must be close to its old host element, the particle is searched next in all elements surrounding the nodes of its old host element. (This is also stored in a linked list for fast access.) This fall-back procedure always finds the particle around a corner, thus a brute-force search is not necessary over all elements.

### 3.7. Particle-number control

In the setup phase an equal number of particles are uniformly generated into each element with the initial velocities $\mathcal{U}_i$ sampled from a Gaussian distribution with zero mean and variance $2/3$, i.e. the initial Reynolds stress tensor is isotropic with unit turbulent kinetic energy, $\langle u_i u_j \rangle = \frac{2}{3}\delta_{ij}$. Initial particle frequencies $\omega$ are sampled from a gamma distribution with unit mean and variance $1/4$ and the scalar concentration $\psi$ is set to 0.

During the timestepping procedure a sufficient number of particles have to be present in every element at all times to keep the deterministic error due to bias small (66). However, the grid can be refined differently in different regions of the domain, as it is done at walls to resolve the boundary layer or around a concentrated source of a passive scalar to capture the high scalar gradients. Since the particles themselves model real fluid particles, at locations where the grid is refined more particles are necessary for an increased resolution. Therefore it is reasonable to keep the element-distribution of the number of particles as homogeneous as possible. Particle-number control is a delicate procedure in PDF methods, because external modification of the particle locations or properties may result in undesired changes of the local statistics and the joint PDF itself. Nevertheless, particle splitting and merging techniques are routinely applied to keep the particle distribution reasonable and to improve the efficiency and stability of the simulation (67). Section 7 describes the algorithm that we developed to keep the number of particles per element above a certain treshold and to guard the simulation against the occurrence of empty elements (i.e. elements without particles).

In what follows, we describe a simple testcase to investigate the error introduced by the particle redisitribution. Note that the traditional way of referring to this procedure is *particle splitting and merging*. Since we do not change the total number of particles throughout the simulation (which is more memory efficient than splitting and merging) we refer to this procedure as *particle redistribution*. To investigate the error, we consider the simplified model equations

$$d\mathcal{X}_i = \mathcal{U}_i dt, \tag{46}$$

$$d\mathcal{U}_i = -(\mathcal{U}_i - \alpha \langle U_i \rangle)dt + \sqrt{2}dW_i, \tag{47}$$

where $\alpha$ is a scalar parameter and the initial conditions for $\mathcal{U}_i$ are taken to be independent, standardized, normally distributed random variables:

$$\langle U_i \rangle = 0, \qquad \langle u_i u_j \rangle = \delta_{ij}. \tag{48}$$

Eq. (47) is characteristic of the Langevin equation (10) without viscous effects, see also (68). The mean $\langle U_i \rangle$ of the solution of the stochastic differential equation (47) is the solution of the following linear deterministic differential equation (69)

$$\frac{d\langle U_i \rangle}{dt} = -(\langle U_i \rangle - \alpha \langle U_i \rangle), \tag{49}$$

$$\langle U_i \rangle(t=0) = 0. \tag{50}$$

It can be seen that the trivial solution $\langle U_i \rangle = 0$ satisfies the above deterministic initial value problem. For a nonzero initial condition the solution of Eq. (47) is stable and reaches steady state if $\alpha < 1$ with $\langle U_i \rangle = 0$ and $\langle u_i u_j \rangle = \delta_{ij}$. For $\alpha > 1$ the equation becomes unstable and the solution grows exponentially, while for $\alpha = 0$ the equation is neutrally stable. For our purposes we use $\alpha = 0.5$. Eqs. (46) and (47) are advanced on a rectangular domain with two free slip walls (from where particles are simply reflected) and a periodic inflow/outflow boundary-pair, see Figure 3. The domain is highly stretched on purpose in the $y$ direction. Initially, an equal number of particles are generated into every element, which in the current case results in a spatially inhomogeneous particle distribution. As the timestepping advances the particles naturally tend to evolve into a spatially homogeneous distribution, which may result in empty elements in the highly refined region if the number of particles is too small. This is circumvented by the particle-redistribution algorithm. We will test the algorithm by calculating the time-evolutions of the spatial average of the diagonal components of $\langle u_i u_j \rangle$, indicated by $\overline{\langle u_i u_j \rangle}$, using different initial number of particles per element $N_{p/e}$. In order to ensure that the particle-redistribution algorithm intervenes on the same level in each case, the ratio

$$\frac{N_{p/e}}{N_{p/e}^{\min}} \propto \text{number of particles moved} \tag{51}$$

is kept constant. In other words, as the initial number of particles $N_{p/e}$ is increased, we increase the required minimum number of particles per element $N_{p/e}^{\min}$ as well, so that the number of particles that will have to be moved is approximately the same, hence the algorithm intervenes at the same level. To verify that this is the case, the number of times the redistribution algorithm is called (the number of particles moved in a timestep) is monitored and plotted in Figure 5 for the different cases. Figure 4 depicts $\overline{\langle u_i u_j \rangle}$ for different values of $N_{p/e}$.

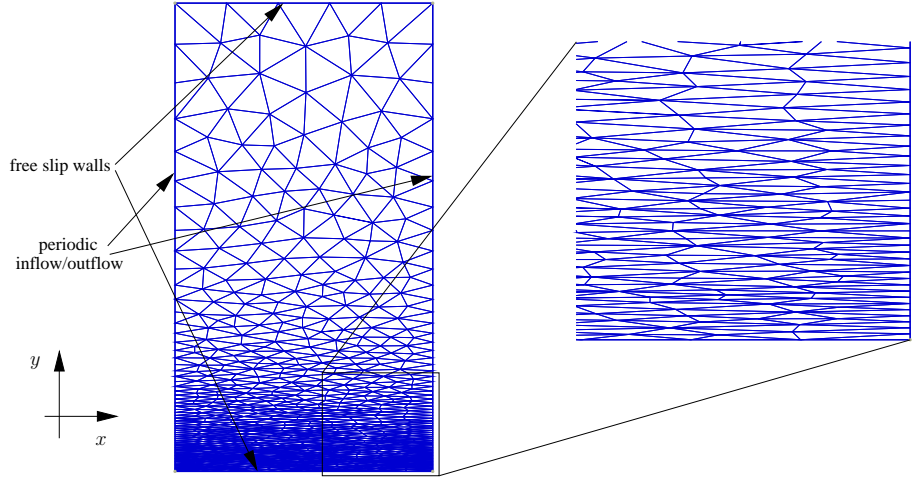

Fig. 3. A rectangular domain with a stretched grid to test the error introduced by the particle-redistribution algorithm using Eqs. (46) and (47).

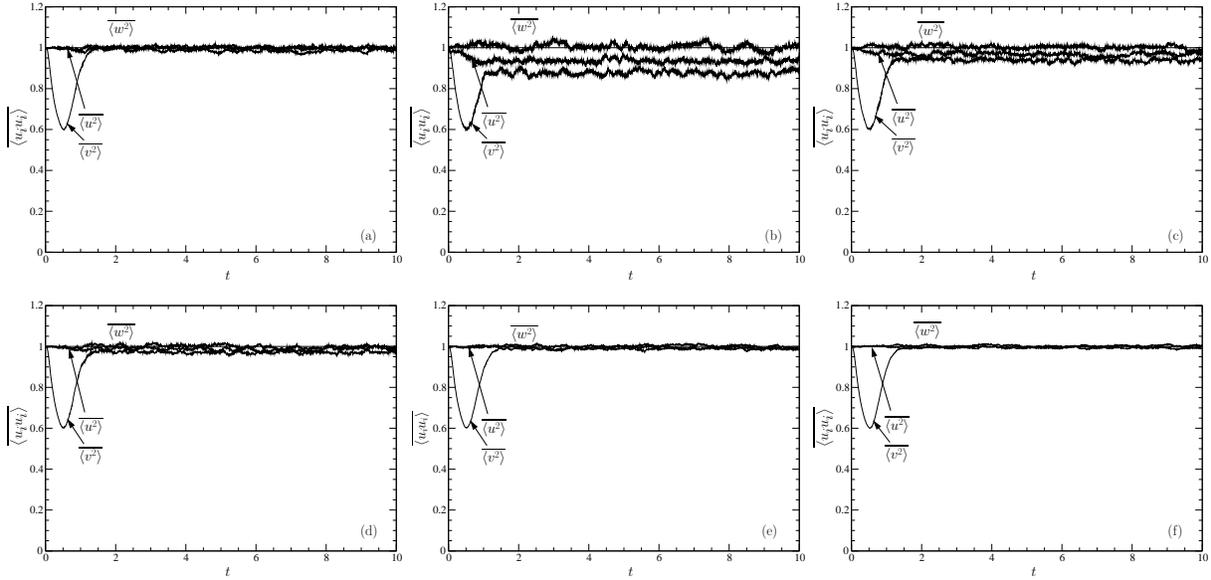

Fig. 4. Time-evolutions of the diagonal components of $\overline{\langle u_i u_j \rangle}$ solving Eqs. (46) and (47) employing different number of particles. (a) No redistribution with initial number of particles per element $N_{p/e}$=200; redistribution with (b) $N_{p/e}$=50, (c) $N_{p/e}$=100, (d) $N_{p/e}$=200, (e) $N_{p/e}$=400 and (f) $N_{p/e}$=800, respectively. The ratio $N_{p/e}/N_{p/e}^{\min}$=10 is kept constant for cases (b) to (f). The horizontal line at the ordinate 1 depicts the analitical solution at steady state.

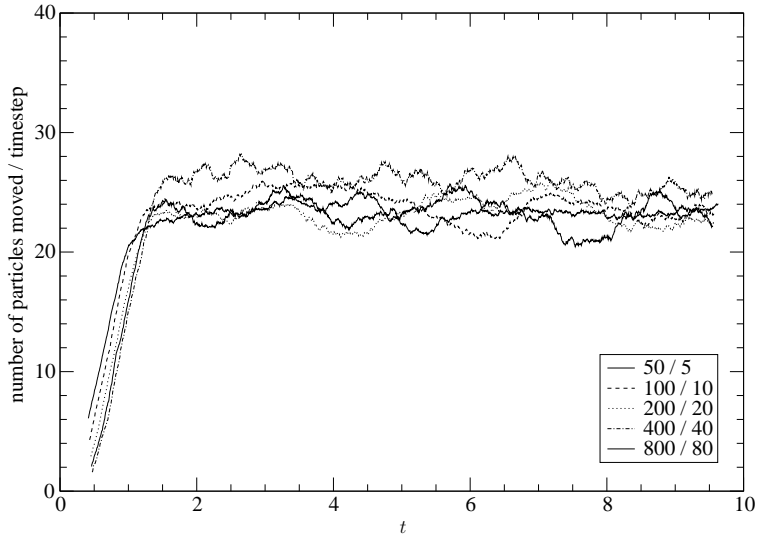

Fig. 5. The number of particles moved in each timestep by the particle-redistribution algorithm for different total number of particles. In the legend the constant $N_{p/e}/N_{p/e}^{\min}$ ratio is displayed.

It can be seen in Figure 4 (a) that the algorithm reproduces the analitical solution with a given numerical error. This error, which is always present in the numerical solution of stochastic differential equations, can be decomposed into three different parts: truncation error due to finite-size timesteps, deterministic error (or bias) due to the finite number of particles employed and random (or statistical) error (66). The particle redistribution introduces an additional error which is directly visible by comparing Figures 4 (a) and (d). It is also apparent that the bias decreases with increasing number of particles as it can be expected. However, Figures 4 (b)-(f) also show that the additional error introduced by the particle redistribution also diminishes as the number of particles increase while the intervention of the redistribution, Eq. (51), is kept at a constant level. This can be seen more directly in Figure 6, which depicts the evolution of the total relative numerical error defined as

$$\delta = \frac{k_c - k_a}{k_a}, \tag{52}$$

where $k_c$ and $k_a$ denote the computed and analytical turbulent kinetic energy, respectively. This error incorporates both the usual numerical errors and the additional one due to the particle-redistribution algorithm. For comparison, the evolution of the error without particle redistribution is also displayed. Since the total sum of the errors converges to zero, the error introduced by the redistribution algorithm also diminishes and the solution converges to the PDF without redistribution.

We have found that a particle-redistribution algorithm of a similar sort (or particle splitting and merging) is essential to provide adequate numerical stability in modeling inhomogeneous flows especially in complex geometries. In addition, it also dramatically reduces the need for high number of particles elements on stretched grids.

### 3.8. Wall-boundary conditions

Over any given time-interval a particle undergoing reflected Brownian motion in the vicinity of a wall may strike the wall infinitely many times (41). This means that particles can follow three different trajectories when interacting with walls. The particle either (a) crosses the wall during the timestep and it is behind the wall at the end of the timestep or (b) crosses the wall during the timestep but it is not behind the wall at the end of the timestep or (c) does not cross the wall during the timestep. Therefore wall-conditions have to be enforced on particles that follow trajectory (a) and (b). The probability that the particle following trajectory (b) crossed the wall during timestep $\Delta t$ can be calculated by (70)

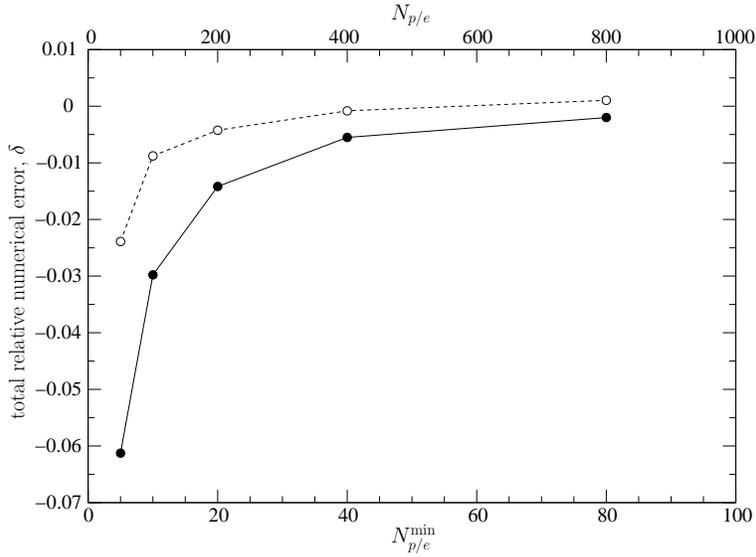

Fig. 6. Evolution of the total relative numerical error defined by Eq. (52) with increasing number of particles. Solid line – with redistribution, dashed line – without redistribution.

$$f_w = \exp\left(-\frac{d^n d^{n+1}}{\nu \Delta t}\right), \tag{53}$$

where $d^n$ denotes the distance of the particle from the wall at timestep $n$. Thus, particle wall-conditions are applied if

$$d^{n+1} < 0, \quad \text{trajectory (a)}, \tag{54}$$

or if

$$d^{n+1} \geq 0 \quad \text{and} \quad \eta < f_w, \quad \text{trajectory (b)}, \tag{55}$$

where $\eta$ is a random variable with a standard uniform distribution. The new particle location is calculated based on perfect reflection from the wall, the particle velocity is set according to the no-slip condition

$$\mathcal{U}_i = 0. \tag{56}$$

A boundary condition on the characteristic turbulent frequency $\omega$ has to ensure the correct balance of the turbulent kinetic energy at the wall (41) and has to be consistent with the near-wall kinetic energy equation

$$\nu \frac{\partial^2 k}{\partial n^2} + \varepsilon = 0, \tag{57}$$

where $\boldsymbol{n}$ is the outward normal of the wall. Accordingly, the particle frequency for a particle striking the wall is sampled from a gamma distribution with mean and variance respectively (41)

$$\langle \omega \rangle = \frac{1}{C_T} \frac{\mathrm{d}\sqrt{2k}}{\mathrm{d}y} \quad \text{and} \quad \left\langle \left(\omega - \langle \omega \rangle\right)^2 \right\rangle = C_4 \langle \omega \rangle^2. \tag{58}$$

For better performance the above particle conditions are only tested and enforced for particles that reside close to walls, i.e. elements that have at least an edge or a node on a no-slip wall-boundary.

Following (41), the wall-boundary condition for the elliptic relaxation equation (12) is set according to

$$\wp_{ij} = -4.5\varepsilon n_i n_j. \tag{59}$$

For the pressure-Poisson equation (33), a Neumann-condition is obtained from the Eulerian mean-momentum equation

$$\frac{\partial \langle U_i \rangle}{\partial t} + \langle U_j \rangle \frac{\partial \langle U_i \rangle}{\partial x_j} + \frac{1}{\rho} \frac{\partial \langle P \rangle}{\partial x_i} = \nu \nabla^2 \langle U_i \rangle - \frac{\partial \langle u_i u_j \rangle}{\partial x_j}, \tag{60}$$

by taking the normal component at a stationary solid wall

$$\frac{1}{\rho}\frac{\partial \langle P \rangle}{\partial x_i}n_i = \nu \frac{\partial^2 \langle U_i \rangle}{\partial x_j \partial x_j}n_i - \frac{\partial \langle u_i u_j \rangle}{\partial x_j}n_i. \qquad (61)$$

### 3.9. Parallel random number generation

The solver has been parallelized and run on different shared memory architectures. Both the initialization and the timestepping require a large number of random numbers with different distributions and characteristics. Two components of the position $\mathcal{X}_i$ and three components of the velocity $\mathcal{U}_i$ are retained for a two-dimensional simulation, therefore the governing equations (7), (10) and (19) altogether require 6 independent Gaussian random numbers for each particle in each timestep. Since these 6 numbers per particle are always needed and are always Gaussian, they can be efficiently stored in a table, which is regenerated in each timestep. Different methods exist to efficiently sample pseudo-random numbers in parallel (71). In order to be able to reproduce the simulation results and to avoid surpassing cross-correlations between random number streams, we initialize a single stream and split it into $k$ non-overlapping blocks, where $k$ is the number of parallel threads. Then each of the threads generates from its own corresponding block, avoiding data races with other threads. This can be quite efficient, since a large amount of random numbers are generated at once and each thread accesses only its own portion of the stream. The same block-splitting technique is used to fill another table with uniform random numbers for the boundary condition Eq. (55). Using this sampling technique, an almost ideal speedup can be achieved when random numbers in tables are regenerated, see also Table 1. For those equations in which the number of random numbers is a priori unknown (e.g. sampling a gamma distribution for the wall-condition of Eq. (58) for particles that struck the wall), a stream is split into $k$ disjoint substreams and the leap-frog technique is used to sample from them in parallel (72). These techniques have been found essential to achieve a good parallel performance for the loop advancing the particles, see also Section 3.10.

### 3.10. Solution procedure and execution profile

The main stages of one complete timestep in their order of execution are displayed in Table 1. Also shown are the percentage of the execution times of each stage relative to a complete timestep and their speedups on a machine with two quad-core processors. The performance data were obtained by running a case that contained approximately 10 million particles and the Eulerian grid consisted of about 20 thousand triangles.

Table 1
Structure and profile of a timestep with relative execution times compared to the time spent on the full timestep and parallel performances of each step on a machine with two quad-core processors. The listing order corresponds to the order of execution. The performance data is characteristic of a case with 10M particles using a grid with 20K triangles, the simulation altogether requiring approximately 1.2GB memory. The processors are two quad-core CPUs (8 cores total), each pair sharing 4MB cache and the CPU-to-memory communication bandwidth.

| task | relative execution time | speedup with 2 CPUs | speedup with 4 CPUs | speedup with 6 CPUs | speedup with 8 CPUs |
|---|---|---|---|---|---|
| compute the size of the next timestep, see Section 3.1 | 0.001 % | not parallelized | | | |
| solve elliptic relaxation equation (12), see Section 3.2 | 2.87 % | 1.91 | 4.08 | 5.76 | 7.60 |
| advance particle properties according to Eqs. (7), (10), (19) and (24) | 73.2 % | 2.02 | 4.12 | 6.16 | 8.20 |
| regenerate random number tables, see Section 3.9 | 19.01 % | 2.01 | 3.99 | 5.79 | 7.50 |
| solve pressure-Poisson equation, see Section 3.2 | 2.0 % | 1.86 | 3.49 | 4.55 | 5.02 |
| correct mean velocities, see Section 3.2 | 1.0 % | 1.69 | 1.95 | 1.94 | 1.96 |
| compute Eulerian statistics, see Sections 3.3–3.5 | 1.6 % | 1.22 | 1.79 | 1.67 | 1.77 |
| one complete timestep | 99.68 % | 1.98 | 3.95 | 5.55 | 7.20 |

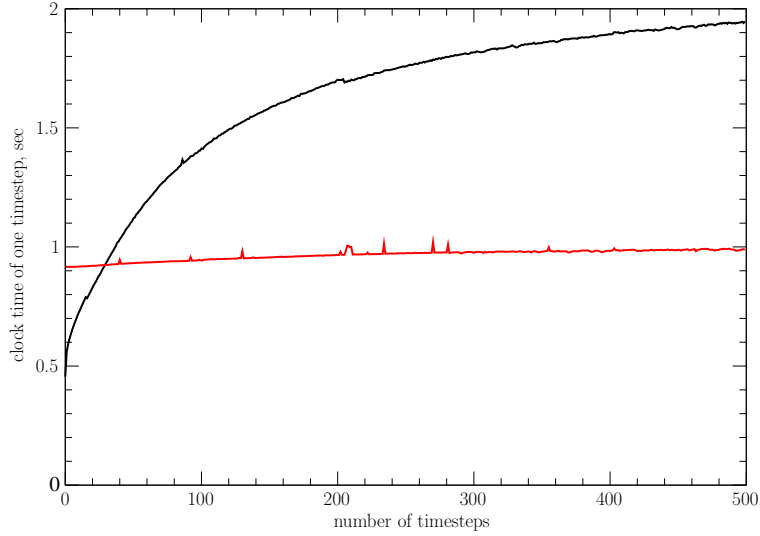

Fig. 7. Performance comparison of the two different loops (displayed in Table 2) to advance the particle governing equations (7), (10), (19) and (24) for the first 500 timesteps using 8 CPUs. The almost horizontal (red) line represents the particle-based loop, while the curving (black) one is the element-based loop. The problem size is the same as in Table 1.

A significant portion of the execution time is spent on advancing the particle-governing equations. This is mostly a loop which can be constructed in two fundamental ways: in an element-based or in a particle-based fashion as displayed in Table 2. The main advantage of the element-based loop is that once the Eulerian statistics are gathered for an element they can be used to update all particles in the element without recomputing them. However, it can be significantly off-balance in parallel, since it is not rare that the number of particles per element can differ by as much as two orders of magnitude at different regions of the domain. Another disadvantage of the element-based loop is that most of the time it accesses the arrays containing the particle properties, $\mathcal{X}_i$, $\mathcal{U}_i$, $\omega$, $\psi$, in an unordered fashion resulting in increasing cache misses as the timestepping progresses and the particles move throughout the domain, because they get scrambled in memory compared to their spatial locations. Conversely, the big advantages of the particle-based loop is its simplicity and excellent load-balance for parallel execution, although it has to gather and recompute the statistics for each particle, including the particles residing in the same element whose statistics have already been computed. The particle-based loop always accesses the arrays containing particle properties consecutively. The effect of the increasing cache misses and the different load-balance on the performance is displayed in Figure 7, where the timings of the two loops are compared as the iteration progresses. The element-based loop slows down almost fourfold in just 500 timesteps, while the performance degradation

Table 2
Two fundamental ways of constructing a loop to advance the particle-governing equations (7), (10), (19) and (24). Left – element-based loop, right – particle-based loop.

```
for all Eulerian elements e                              for all particles p

    gather Eulerian nodal statistics for element e;          obtain index e of host element for particle p;
    compute element-average statistics;                      gather Eulerian nodal statistics for element e;
                                                             compute element-average statistics;
    for all particles p in element e                         advance particle p;
        advance particle p;
    end                                                  end

end
```

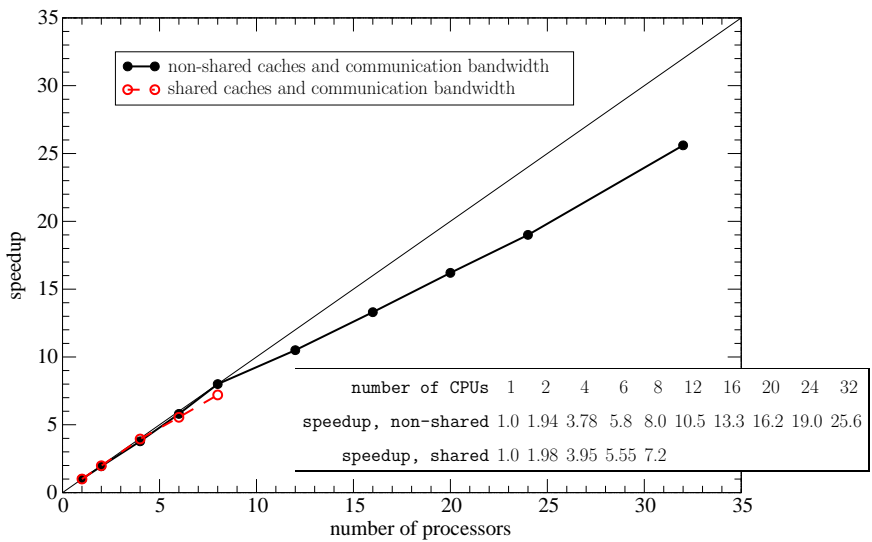

Fig. 8. Overall parallel performance of 100 timesteps taken on two different types of shared memory machines. Solid line and symbols – separate caches and memory-to-CPU bandwidths for each processor, dashed line and open symbols – two quad-core CPUs (8 cores total) each pair sharing a cache and a memory-to-CPU bandwidth. The problem size is the same as in Table 1.

of the particle-based loop is negligible. Also, this disparity increases as the number of threads increases, which is shown in Table 3, where serial and parallel timings are displayed for both loops with different number of threads. While the element-based loop slightly outperforms the particle-based loop on a single CPU, the high scalability and cache-efficiency of the particle-based loop pays out very well in parallel. In fact its speedup is superlinear, which is due to the fact that as the number of processors increase, more and more data gathered from memory fit into the aggregate cache of the individual CPUs, resulting in faster processing than from central memory.

Cache misses may also be reduced by specifically optimizing for the architecture of shared caches on multicore CPUs as it has been done in the current case. We have found that this guarantees a good performance on true shared memory machines as well, i.e. on machines whose CPUs do not share their caches and the communication bandwith between the CPU and memory. However, optimizing for non-shared caches and communication bandwidths does not necessarily guarantee optimal performance on multicore CPUs. These findings clearly show the importance of efficient use of caches. This was also noted with Eulerian CFD codes computing a variety of flows (73).

The parallel performance on higher number of processors is plotted in Figure 8. The size of the testproblem is the same as in Table 1, but the hardware is now a true shared memory machine with separate cache and memory-to-CPU bandwidth for each processor. The code performs reasonably well for this moderate-size problem and the parallel efficiency does not show a sign of leveling out up to the 32 CPUs tested. For comparison, the performance data in Table 1 is also shown using mutlicore CPUs.

Table 3
A comparison of serial and parallel performances for a single timestep of the most time-consuming loop, implementing the governing equations to advance particles, Eqs. (7), (10), (19) and (24), using the two different loop-structures displayed in Table 2. The data is obtained from the same test simulation as in Table 1 using the same hardware. The timings are approximate values after the first 500 timesteps.

| | element-based loop | | particle-based loop | |
| number of CPUs | time (ms) | speedup | time (ms) | speedup |
| --- | --- | --- | --- | --- |
| 1 | 6909 | 1.0 | 8068 | 1.00 |
| 2 | 4122 | 1.68 | 3987 | 2.03 |
| 4 | 2408 | 2.87 | 1943 | 4.12 |
| 6 | 1979 | 3.49 | 1305 | 6.16 |
| 8 | 1945 | 3.55 | 1000 | 8.20 |

Table 1 shows that the second most time-consuming step in a timestep is the regeneration of the random number tables, which was discussed in Secion 3.9. Interestingly, the solution of the two Eulerian equations, namely the elliptic relaxation equation (12) and the pressure-Poisson equation (33), only take up about 2-3% of a timestep, respectively. It is worth noting that the linear system for the elliptic relaxation is nine times larger than that of the pressure-Poisson equation. The former is very well conditioned, while the latter is usually the most time-consuming equation to solve in modeling laminar incompressible flows.

## 4. Testcases

Two testcases demonstrate the applicability of the algorithm: a fully developed turbulent channel flow and a street canyon (or cavity) flow both with passive scalar releases from a concentrated source. For validation, several statistics are compared to direct numerical simulation (DNS) and experimental data. The effects of various numerical parameters on the results, such as the number of conditioning bins in the estimation of $\langle \phi | \boldsymbol{V} \rangle$ or the number of particles, are also analyzed in the next subsections.

### 4.1. *Scalar dispersion in fully developed turbulent channel flow*

The velocity field in turbulent channel flow, after an initial development time, becomes statistically stationary and homogeneous in the streamwise direction, while it remains inhomogeneous in the wall-normal direction, i.e. the flow becomes statistically one-dimensional. The flow is assumed to be statistically symmetric about the channel centerline. A passive scalar released into this flow is inhomogeneous and three-dimensional. Assuming the channel cross section has a high aspect ratio, we confine our interest to the plane spanned by the wall-normal and streamwise directions, far from the spanwise walls. The computational scheme exploits these features by resolving only one spatial dimension for the velocity statistics and two dimensions for the passive scalar. Although, this specialized implementation of the method includes flow-dependent features, it provides good indication of the total computational cost. The description is divided into sections that separately discuss the modeling of the fluid dynamics (Section 4.1.1) and the transported scalar (Section 4.1.2). Both DNS and experimental data are used to validate the results.

#### 4.1.1. *Modeling the fluid dynamics*

Since the transported scalar is inhomogeneous, both streamwise $x$ and cross-stream $y$ components of the particle positions are retained. A one-dimensional grid is used to compute Eulerian statistics of the velocity and turbulent frequency. An increasing level of refinement is achieved in the vicinity of the wall by obtaining the spacing of the gridpoints from the relation

$$y^+ = 1 - \cos\left(\frac{\pi}{2}a^{3/4}\right), \qquad 0 \leq a < 1, \tag{62}$$

where $y^+ = u_\tau y/\nu$ is the distance from the wall non-dimensionalized by the friction velocity $u_\tau$ and the kinematic viscosity $\nu$ and $a$ is a loop-variable that equidistantly divides the interval between 0 and 1 (wall and centerline, respectively) into a desired number of gridpoints. The centerline symmetry of the flow is exploited, thus these statistics are only computed on half of the channel. Using this one-dimensional grid, Eulerian statistics are computed as described in Section 3.3. First and second derivatives of the mean velocity are calculated by first-order accurate finite difference formulas over each element and then transferred to nodes. A constant unit mean streamwise pressure gradient is imposed, which drives the flow and builds up the numerical solution. The cross-stream mean-pressure gradient is obtained by satisfying the cross-stream mean-momentum equation for turbulent channel flow

$$\frac{1}{\rho}\frac{\mathrm{d}\langle P \rangle}{\mathrm{d}y} = -\frac{\mathrm{d}\langle v^2 \rangle}{\mathrm{d}y}, \tag{63}$$

Table 4
Constants for modeling the joint PDF of velocity and frequency.

| $C_1$ | $C_2$ | $C_3$ | $C_4$ | $C_T$ | $C_L$ | $C_\eta$ | $C_v$ | $\gamma_5$ | $C_{\omega 1}$ | $C_{\omega 2}$ |
|---|---|---|---|---|---|---|---|---|---|---|
| 1.85 | 0.63 | 5.0 | 0.25 | 6.0 | 0.134 | 72.0 | 1.4 | 0.1 | 0.5 | 0.73 |

which implies that the pressure-projection is not necessary for this flow. Since the number of elements do not exceed 100, particle tracking in this one-dimensional case is simply a brute-force check on each element. This is a negligible fraction of the running time, thus there is no need for a more sophisiticated tracking algorithm.

Wall-boundary conditions for the particles are the same as described in Section 3.8, only the situation is simpler here, since the wall is aligned with the coordinate line $y = 0$. The conditions for the centerline are symmetry conditions, i.e. particles trying to leave the domain through the centerline undergo perfect reflection and the sign of their wall-normal velocity is reversed. Consistently with these particle conditions, boundary conditions are imposed on the Eulerian statistics as well. At the wall, the mean velocity and the Reynolds stress tensor is forced to zero. The mean frequency $\langle\omega\rangle$ is set according to Eq. (58). At the centerline, the shear Reynolds stress $\langle uv\rangle$ is set to zero. At the wall in the elliptic-relaxation equation (12), $\wp_{ij}$ is set according to $\wp_{ij} = -4.5\varepsilon n_i n_j$. In the current case the wall is aligned with $y = 0$ thus only the wall-normal component is non-zero: $\wp_{22} = -4.5\varepsilon$. At the centerline, symmetry conditions are enforced on $\wp_{ij}$, i.e. homogeneous Dirichlet-conditions are applied for the off-diagonal components and homogeneous Neumann-conditions for the diagonal components. The initial conditions for the particles are set according to Section 3.7, however the current one-dimensional case enables the use of a sufficient number of particles so that there is no need for particle redistribution. The applied model constants for the joint PDF of velocity and frequency are displayed in Table 4.

### 4.1.2. Modeling the passive scalar

A passive, inert scalar is released from a concentrated source into the modeled fully developed turbulent channel flow, described above. Since the scalar field is inhomogeneous and, in general, not symmetric about the channel centerline, a second, two-dimensional grid is employed to calculate scalar statistics. Employing separate grids for the fluid dynamics and scalar fields enables the grid refinement to be concentrated on different parts of the domain, i.e. the scalar-grid can be refined around the source, while the fluid dynamics-grid is refined at the wall. The two-dimensional mesh is used to calculate Eulerian scalar statistics as described in Section 3.3. Since the scalar statistics are not homogeneous in the streamwise direction, the long rectangular domain is subdivided into several bins (thin vertical stripes, see Figure 9) and the following strategy is used to exploit these features. The velocity and turbulent frequency statistics are computed using the one-dimensional grid in which only particles in the first bin participate. The position of these particles are then copied to all downstream bins and (since the fluid dynamics is symmetric about the channel centerline) these particle positions are also mirrored to the upper half of the channel. This means that the particles (as far as positions are concerned) never leave the first bin physically. Since the scalar is passive, only one-way coupling between the two grids is necessary. This is accomplished by using the local velocity statistics computed in the 1d-elements for those 2d-elements that lie the closest to them in the wall-normal coordinate direction. At the wall and centerline boundaries the conditions on the particle properties have already been described in Section 4.1.1. For particles trying to leave the bin through the "inflow/outflow" bin-boundaries a periodic boundary condition is applied, with leaving particles put back on the opposite side. This essentially means that the particle paths remain continuous (as they should), only the code accounts for them as different particles in the computer memory. In order to carry the scalar concentration through bin-boundaries, the particle-scalar $\psi$ is copied downstream (upstream) when the particle tries to leave through the downstream (upstream) bin-boundary. If the particle hits the centerline, its concentration is exchanged with its mirrored pair on the upper half, facilitating a possible non-symmetric behaviour of the scalar. The line-source, which in the current two-dimensional case is a point-source, is represented by a circular source with diameter 0.05. The scalar at the source has a constant distribution: particles passing through the source are assigned a constant normalized unit source strength, i.e. $\psi = \phi_0 = 1$. The applied model constants for the micromixing timescale defined by Eq. (25) are $C_s = 0.02$ and $C_t = 0.7$.

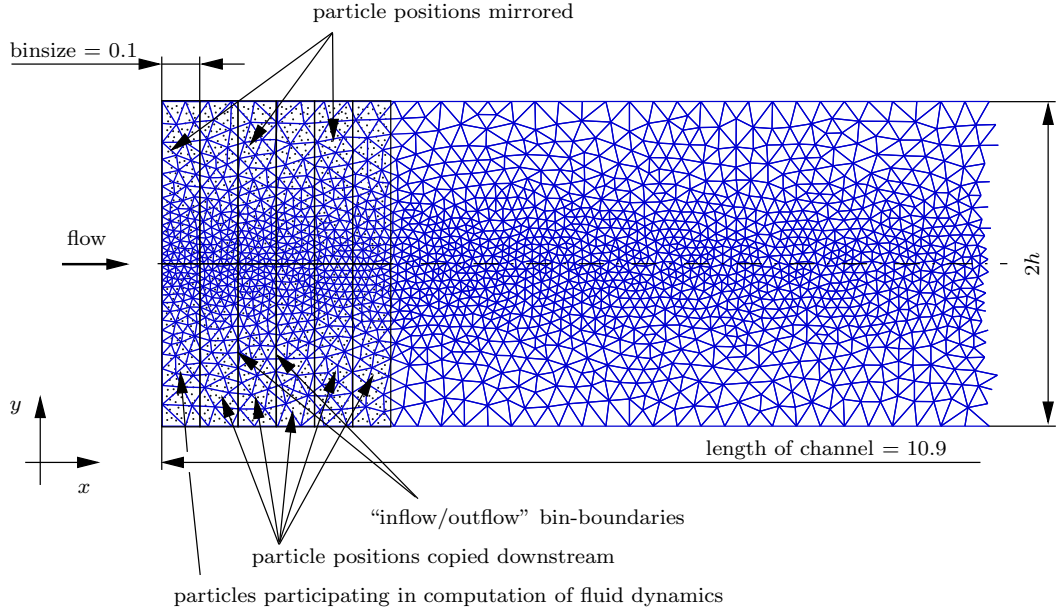

Fig. 9. The computational domain for the channel flow is subdivided into several bins to exploit the streamwise statistical homogeneity of the turbulent velocity and frequency fields. Particle positions are copied downstream and mirrored to the upper half. Particle scalar concentrations are exchanged through bin-boundaries and the centerline. Note, that the number of particles in the figure does not correspond to the actual number used in the simulation.

4.1.3. *Results for channel flow*

Previous PDF modeling studies of channel flow in conjunction with elliptic relaxation have been reported at $Re_\tau = 395$ (41) and $Re_\tau = 590$ (77) based on the friction velocity $u_\tau$ and the channel half-width $h$. These works concentrate on model development and employ different methodologies with different model constants and numerical methods, which inevitably result in a different balance of model behavior and numerical errors. To assess the prediction at different Reynolds numbers the current model has been run at $Re_\tau = 392$, 642 and 1080 using the model constants displayed in Table 4. The velocity statistics for all three cases are depicted in Figure 10. The mean velocity is well represented in the viscous sublayer ($y^+ < 5$) for all three Reynolds numbers. In the buffer layer ($5 < y^+ < 30$) there is a slight departure from the DNS data as the Reynolds number increases and from $y^+ > 30$, where the log-law should hold, there exists approximate self-similarity, i.e. the universal slopes of the profiles are equally well-represented with a slight underprediction far from the wall at higher Reynolds numbers. The viscous wall region ($y^+ < 50$) contains the highest turbulent activity, where production, dissipation, turbulent kinetic energy and anisotropy reach their peak values. The location of the peaks of the Reynolds stress components are succesfully captured by the model at all three Reynolds numbers with their intensity slightly underpredicted. Previous studies using elliptic relaxation in the Reynolds stress framework (i.e. Eulerian RANS models) report excellent agreement for these second-order statistics (40; 51). Wacławczyk *et al.* (77) also achieve very good agreement with DNS data using a different version of a PDF model than the one applied here. A common characteristic of PDF models is the slight overprediction of the wall-normal Reynolds stress component $\langle v^2 \rangle$ far from the wall. This component is responsible for the cross-stream mixing of a transported scalar released into a flow far from a wall. Therefore in applications where the mean concentration of scalars is important this quantity must be adequately captured. To improve on this situation we introduced a slight modification into the computation of the characteristic lengthscale $L$ in the elliptic relaxation equation (16) compared to (41), by inserting the parameter $C_\xi$ as described in Section 2. This only affects the diagonal Reynolds stresses which can be seen in Figure 11 for the different Reynolds numbers. Decreasing $\langle v^2 \rangle$ at the centerline changes the relative fraction of energy distributed among the diagonal components of the Reynolds stress tensor, consequently the other two components, $\langle u^2 \rangle$ and $\langle w^2 \rangle$, are slightly increased. Obviously, these kind of flow-dependent modifications in the turbulence model are of limited value, since their effects in a general setting may not be easily predictable. The only nonzero shear stress component $\langle uv \rangle$ in this flow and the turbulent kinetic energy dissipation rate $\varepsilon$ are both in very good agreement with DNS data and even improve as the Reynolds number increases. It is apparent in both Figures 10 and 11 that the overall prediction of second order statistics improve as the Reynolds number increases. This tendency is expected to continue as the underlying high-Reynolds-number modeling assumptions become better fullfilled.

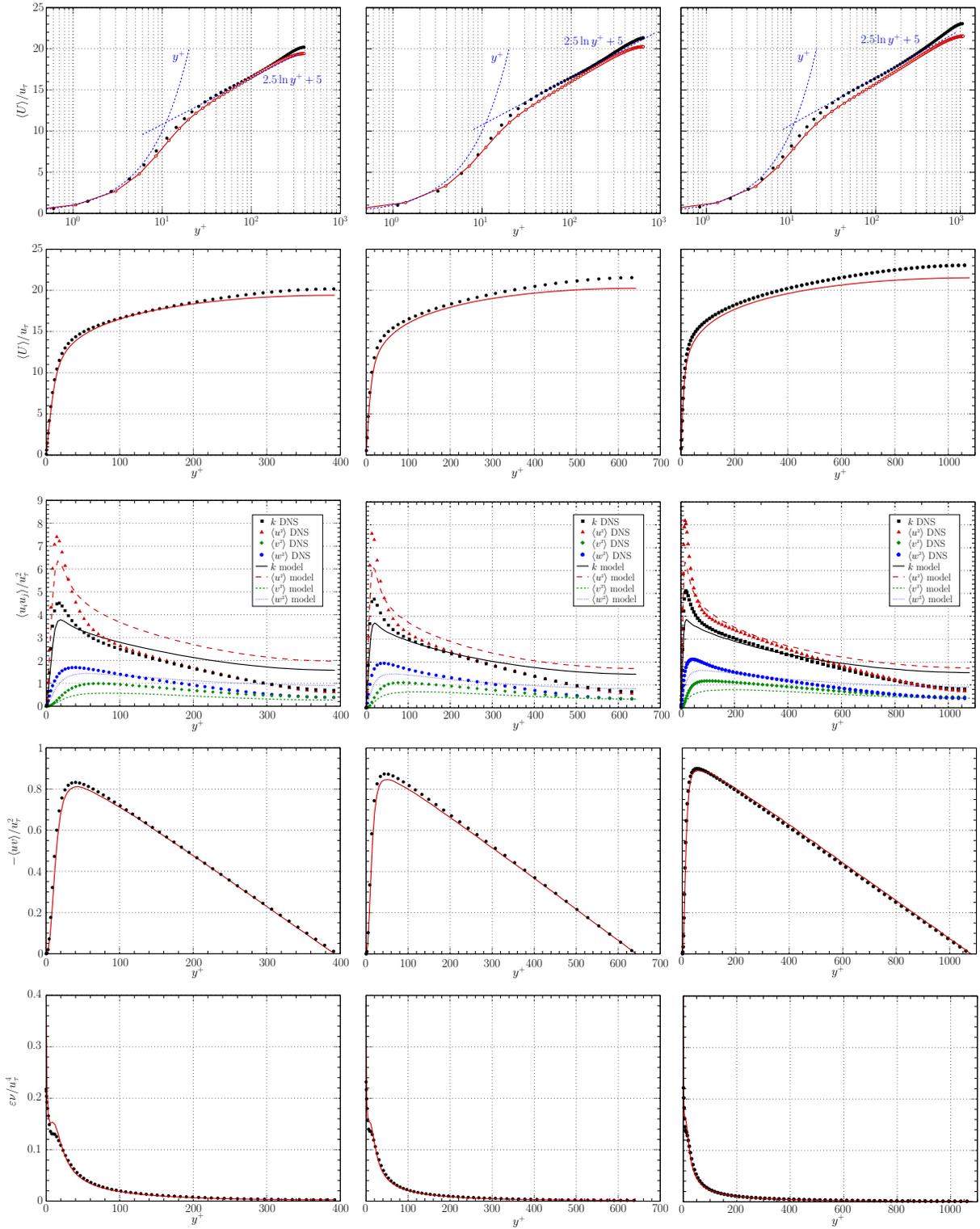

Fig. 10. Cross-stream velocity statistics for fully developed turbulent channel flow at (first column) $Re_\tau = 392$, (middle column) $Re_\tau = 642$ and (right column) $Re_\tau = 1080$. Lines – PDF calculation, symbols – DNS data of Moser et. al (74), Iwamoto *et al.* (75) and Abe *et al.* (76) (scaled from $Re_\tau = 1020$), respectively. First two rows – mean streamwise velocity, third row – normal Reynolds stresses, fourth row – shear Reynolds stress and fifth row – rate of dissipation of turbulent kinetic energy. All quantities are normalized by the firction velocity $u_\tau$ and the channel half-width $h$.

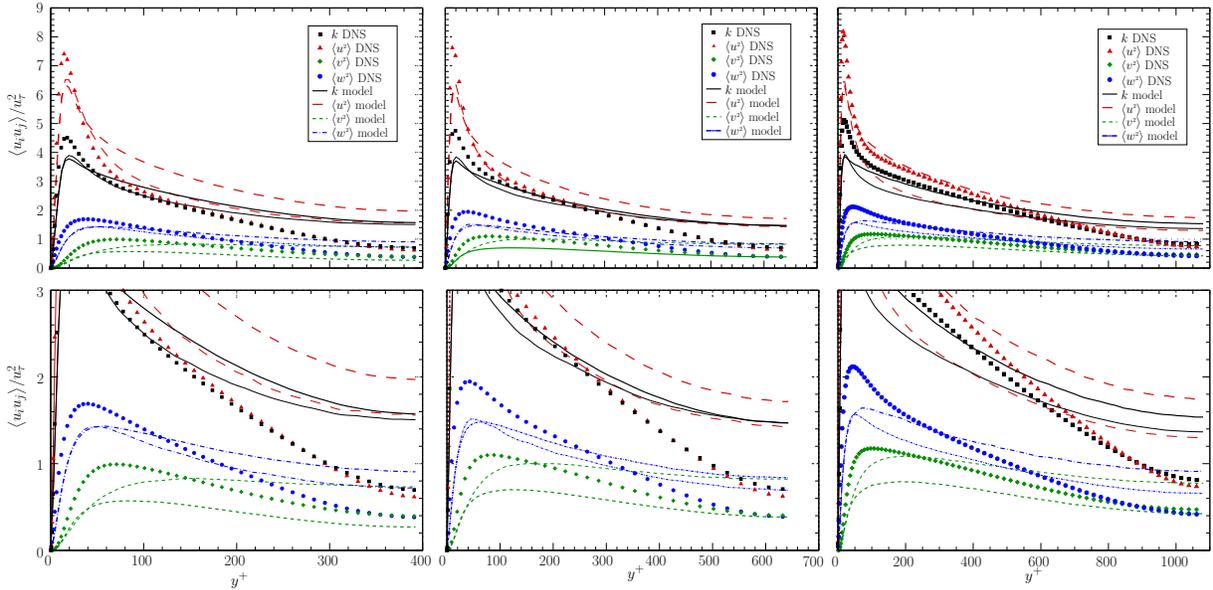

Fig. 11. The effect of the modification of the characteristic lengthscale in Eq. (16) on the diagonal components of the Reynolds stress tensor by employing the additional model constant $C_\xi \neq 1$ at (first column) $Re_\tau = 392$, (middle column) $Re_\tau = 642$ and (right column) $Re_\tau = 1080$. Thick lines, $C_\xi = 1.0 + 1.3 n_i n_i$; thin lines, $C_\xi = 1.0$; symbols, DNS data as in Figure 10. The figures on the bottom row are enlargements of the figures on the top row.

Into the fully developed flow, a passive scalar has been released from a concentrated source at the channel centerline. A general numerical procedure that can be used to compute the velocity-conditioned scalar mean $\langle \phi | \boldsymbol{V} \rangle$ in the IECM model has been described in Section 3.5. Another method based on the projection of the three-dimensional velocity field onto a one-dimensional subspace, where the discretization can be carried out, has been developed and tested in homogeneous turbulence by Fox (56). In that method, the projected velocity of a particle is found from

$$\mathcal{U}_\rho = \alpha_i \mathcal{U}_i, \tag{64}$$

where the projection vector $\alpha_i$ is obtained from the following linear relationship

$$\rho_i = \rho_{ij} \alpha_j \tag{65}$$

between the normalized velocity-scalar vector and the velocity-correlation tensor (no summation on greek indices)

$$\rho_\alpha = \frac{\langle u_\alpha \phi' \rangle}{\langle u_\alpha^2 \rangle^{1/2} \langle \phi'^2 \rangle^{1/2}}, \qquad \rho_{\alpha\beta} = \frac{\langle u_\alpha u_\beta \rangle}{\langle u_\alpha^2 \rangle^{1/2} \langle u_\beta^2 \rangle^{1/2}}, \tag{66}$$

where $\phi' = \psi - \langle \phi \rangle$ denotes the scalar fluctuation. This projection method has been developed (and is exact for) Gaussian velocity PDFs, although it can still be used in inhomogeneous flows with the assumption that the local joint PDF of velocity is not too far from an approximate joint normal distribution. In order to assess the performances and the difference in the predictions, we implemented and compared both methods and tested them with different number of conditioning bins.

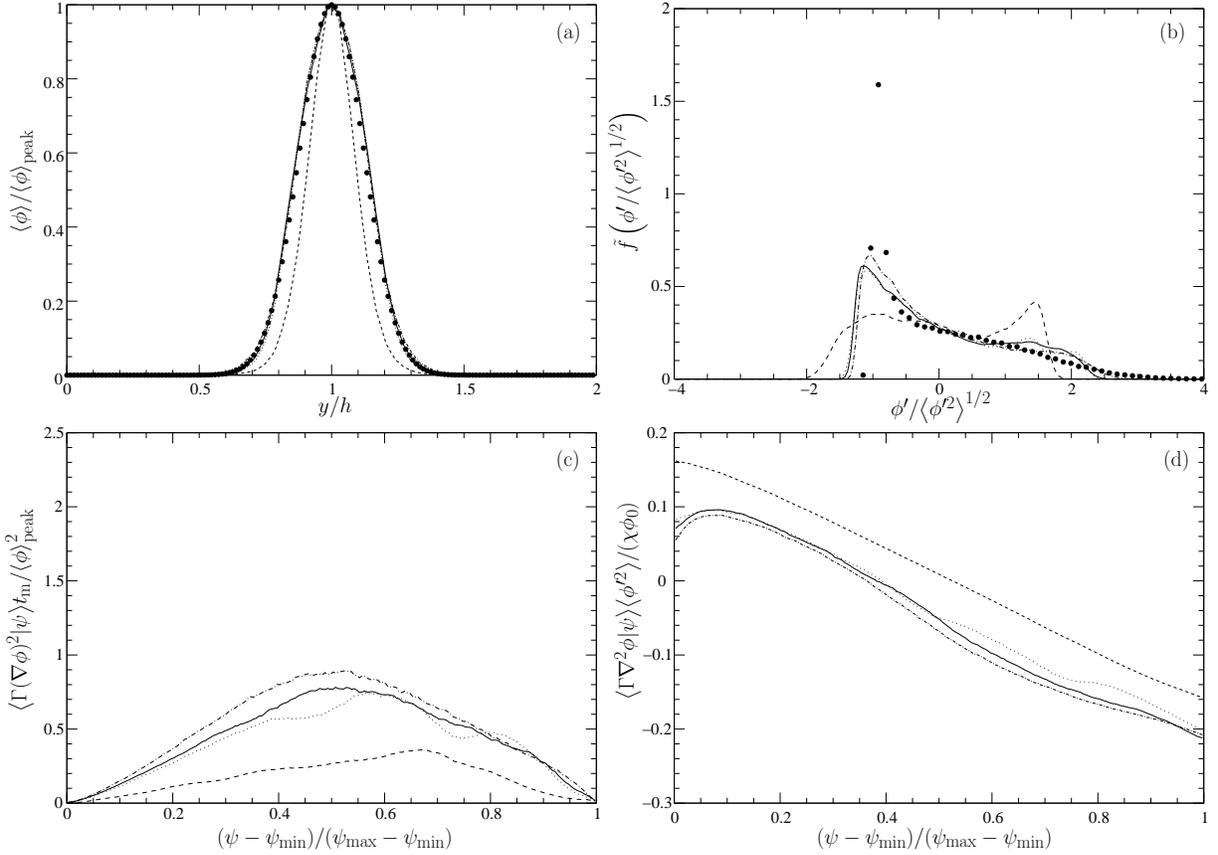

Fig. 12. Scalar statistics affected by the number of conditioning intervals $N_c$ with computing the velocity-conditioned mean $\langle \phi | \boldsymbol{V} \rangle$ applying Fox's projection method using Eqs. (64)-(66). (a) Cross-stream distribution of the scalar mean at $x/h = 4.0$, (b) PDF of scalar concentration fluctuations at $(x/h = 4.0, y/h = 1.0)$, (c) mean scalar dissipation conditioned on the concentration at $(x/h = 4.0, y/h = 1.0)$ and (d) mean scalar diffusion conditioned on the concentration at $(x/h = 4.0, y/h = 1.0)$. Dashed line – $N_c$=1 (IEM), dotted line – $N_c$=3, solid line – $N_c$=5, dot-dashed line – $N_c$=20. Symbols on (a) analytical Gaussians according to Taylor (78) and on (b) experimental data of Lavertu & Mydlarski (79).

To investigate how the choice of the number of conditioning intervals $N_c$ affects the solution with the projection method, several runs have been performed at the highest Reynolds number ($Re_\tau = 1080$) with different values for $N_c$. Some of the unconditional and conditional statistics of the joint PDF are depicted in Figure 12. Note that employing $N_c$=1 corresponds to the special case of the IEM model, Eq. (23). It is apparent that applying only a few intervals already makes a big difference compared to the IEM model in correcting the prediction of the mean concentration and the PDF of concentration fluctuations also moves towards the experimental data. Increasing $N_c$ may be thought as an approach to increase the resolution of the conditioning (thus better exploiting the advantages of the IECM over the IEM model), however, as Fox (56) points out, this is of limited value, since the decreasing number of particles per interval increases the statistical error. The current test simulations have been carried out with an initial 500 particles per element and the total number of particles did not change during simulation. Figure 12 shows that above $N_c$=5 there is no significant change in the statistics and even at $N_c$=20 the results do not deteriorate. Also displayed in Figure 12 are the centerline normalized mean scalar dissipation and diffusion both conditional on the scalar concentration, $\langle \Gamma (\nabla \phi)^2 | \psi \rangle$ and $\langle \Gamma \nabla^2 \phi | \psi \rangle$, respectively. These quantities are important in composition-only PDF methods, where the turbulent velocity field is either assumed or obtained by other means externally. For the IECM model, an exact relationship has been derived by Sawford (80) for the conditional dissipation

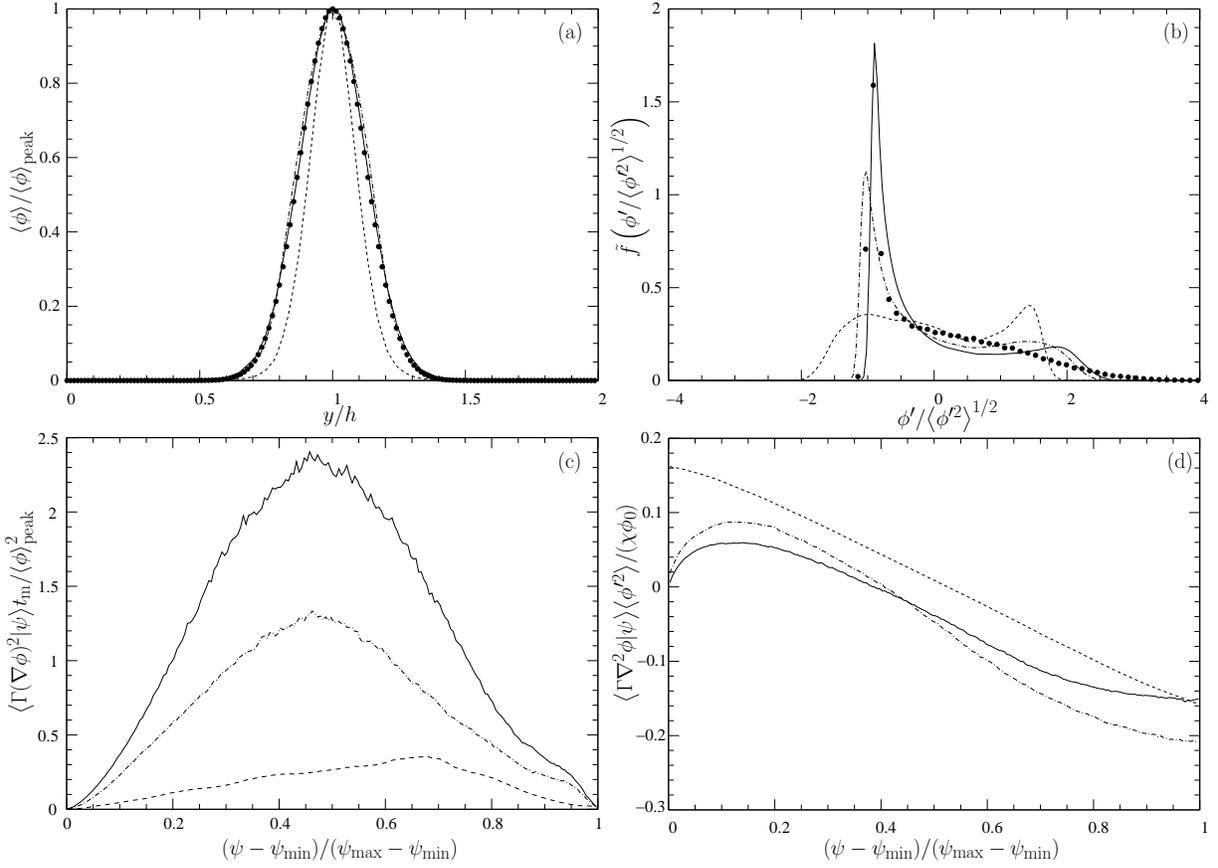

Fig. 13. Scalar statistics affected by the number of conditioning intervals when computing the velocity-conditioned mean $\langle \phi | \boldsymbol{V} \rangle$ with the method described in Section 3.5. The quantities are the same as in Figure 12. Dashed line – $N_c=1$ (IEM), dot-dashed line – $N_c=(3 \times 3 \times 3)$, solid line – $N_c=(5 \times 5 \times 5)$.

$$\left\langle 2\Gamma \frac{\partial \phi}{\partial x_i} \frac{\partial \phi}{\partial x_i} \bigg| \psi \right\rangle \tilde{f} = -\frac{2}{t_m} \int_0^\psi (\psi' - \tilde{\phi}) \tilde{f}(\psi') \mathrm{d}\psi', \tag{67}$$

where

$$\tilde{\phi}(\psi) = \int\int \langle \phi | \boldsymbol{V} \rangle \tilde{f}(\boldsymbol{V}, \omega | \psi) \mathrm{d}\omega \mathrm{d}\boldsymbol{V}, \tag{68}$$

in which the marginal scalar PDF $\tilde{f}(\psi)$ and the conditional PDF $\tilde{f}(\boldsymbol{V}, \omega | \psi)$ are respectively defined based on the full PDF $\tilde{f}(\boldsymbol{V}, \omega, \psi)$ as

$$\tilde{f}(\psi) = \int\int \tilde{f}(\boldsymbol{V}, \omega, \psi) \mathrm{d}\omega \mathrm{d}\boldsymbol{V} \qquad \text{and} \qquad \tilde{f}(\boldsymbol{V}, \omega | \psi) = \tilde{f}(\boldsymbol{V}, \omega, \psi) / \tilde{f}(\psi). \tag{69}$$

Both integrals in Eqs. (67) and (68) can be directly obtained from the simulation. Numerically, the integral in Eq. (68) is obtained by taking the average of $\langle \phi | \boldsymbol{V} \rangle$ over those particles that reside in the bin corresponding to $\psi$. In other words, the concentration values are first conditioned on the velocity field, which is required to advance the particle concentrations according to the IECM model, then are conditioned again by dividing the concentration sample space into bins and computing separate means for each bin. We use a few bins for the velocity conditioning ($N_c$) and a significantly higher number of bins (200) for the scalar sample space in order to obtain a higher resolution, as displayed in Figure 12 (c) and (d). Care must be taken when computing the integral in Eq. (67) numerically, since $\tilde{f}$ may become small at the concentration extremes in the denominator. Special numerical treatments are described in (80) and (57). The conditional diffusion curves are normalized by the scalar variance $\langle \phi'^2 \rangle$, the concentration at the source $\phi_0$ and the mean unconditional dissipation $\chi =$ $\langle 2\Gamma(\nabla\phi)^2\rangle$ which is computed by integrating Eq. (67) over the whole concentration space. The concentration axes in Figure 12 (c) and (d) are scaled between the local minimum and maximum concentration values, $\psi_{\min}$ and $\psi_{\max}$, in order to zoom in on the interesting part of the concentration space. Using Fox's projection method, the choice of number of conditioning intervals on the velocity space ($N_c$) has a similar effect on the conditional dissipation and diffusion: they also support the earlier observation that the optimal number of conditioning intervals is at about $N_c$=3–5 to attain convergence.

A different picture reveals itself however, when $\langle\phi|\boldsymbol{V}\rangle$ is computed with the current method instead of the projection that assumed Gaussianity of the underlying velocity field. The same statistics as shown in Figure 12 are plotted in Figure 13 for different numbers of conditioning bins, but without employing the projection to compute $\langle\phi|\boldsymbol{V}\rangle$. The mean profiles do not behave significantly differently, which underlines the earlier observation that employing only a few conditioning bins can already correct the prediction of the mean compared to the IEM model. The PDFs however show significantly higher spikes when compared to their counterparts with projection. The prediction of the conditional dissipation profiles are also different (overall they range about 150% higher) as opposed to that with projection, while the conditional diffusion curves exhibit similar behavior both with and without projection. Figures 13 (b-d) also reveal that the currently employed finest conditional binning structure of ($5\times5\times5$) with an initial 500 particles element is still not sufficient to achieve convergence for the PDF and these conditional statistics. It is also worth noting that this is the case for a centerline release and that our sampling location is relatively close to the source and at the centerline, which lies in the "approximately homogeneous" region of the flow.

To examine the effect of the number of particles on the solution, several testruns have been performed with different number of particles employing both methods for computing $\langle\phi|\boldsymbol{V}\rangle$. At the Reynolds numbers investigated, $Re_\tau = 392$, 642 and 1080, we found the minimum number of particles per elements necessary for a numerically stable solution to be $N_{p/e}$=80, 100 and 150, respectively. Increasing $N_{p/e}$ more than these minimum values would not be necessary to obtain a particle-number-independent velocity PDF, since running the simulation employing up to $N_{p/e}$=500 resulted in negligible change of the velocity statistics investigated. On the other hand, the scalar statistics exhibit significant differences when different number of particles are employed. Figure 14 shows unconditional and conditional statistics of the passive scalar field at $Re_\tau = 1080$ using different numbers of particles employing the projection method with $N_c$=5. The cross-stream distribution of the first four moments show that the statistical error due to insufficient number of particles becomes higher towards the edge of the plume, where the joint PDF is most skewed. The discrepancy due to this error is more pronounced in the higher-order statistics. The PDFs of concentration fluctuations and the scalar at the centerline, where the flow can be considered approximately homogeneous, is nearly independent of the number of particles. The prediction of accurate conditional statistics usually requires a large number of particles. This is underlined by the mean conditional dissipation and diffusion in Figures 14 (g) and (h) in the center region, which show a slight dependence on $N_{p/e}$. In summary, the velocity statistics are predicted independently of the number of particles. With the projection method to compute $\langle\phi|\boldsymbol{V}\rangle$, the unconditional scalar statistics (including the PDFs) are predicted approximately independently of the number of particles in the homogeneous center region of the channel, however, the conditional statistics examined there still exhibit a slight particle-number-dependence even with $N_{p/e}$=500. We hypothesize that more complex inhomogeneous and highly skewed flows may require even larger number of particles than the currently employed maximum, 500.

In Figure 15 the same scalar statistics as in Figure 14 are shown but with $\langle\phi|\boldsymbol{V}\rangle$ computed with the current method instead of projection for different number of particles employing a binning structure of ($5\times5\times5$). The technique described in Section 3.5 is robust enough to automatically use less conditioning intervals depending on the number of particles in a given element. Thus, when the simulations were run with $N_{p/e}$=150, 300 and 500, the average number of conditioning bins employed throughout the simulation has been automatically reduced to about 57, 100 and 124, respectively, as compared to the prescribed 125. The scalar mean is predicted equally well as with the projection method showing no sign of dependence on the number of particles, Figure 15 (a). Interestingly, the r.m.s. curves do not double-peak if the projection is not used, Figure 15 (b) and the width also agrees better with the experimental data. Thus the double-peaks on Figure 14 (b) may only be artifacts of the projection. Similarly to using projection, the skewness and kurtosis profiles are predicted with significant particle-number dependence at the edges of the plume. This shows that convergence has not yet been reached with $N_{p/e}$=500 for these higher-order statistics. Also, there is a pronounced flattening at the centerline in the skewness and kurtosis profiles using the projection technique, cf. Figures 14 and 15 (c-d), which may also be a side-effect of the projection, since no flattening can be observed in the experimental data. The increasing peaks of the PDFs have already been observed before, when we compared the projection method to the general methodology using different values of $N_c$. Both Figure 15 (e) and (f) show that the PDFs have not converged yet, however, these figures may show the combined effect of increasing both $N_{p/e}$ and $N_c$, since the conditioning algorithm automatically reduces $N_c$ in case of insufficient number of particles in an Eulerian element. Finally, the conditional dissipation and diffusion curves show a very light dependence on the number of particles applied.

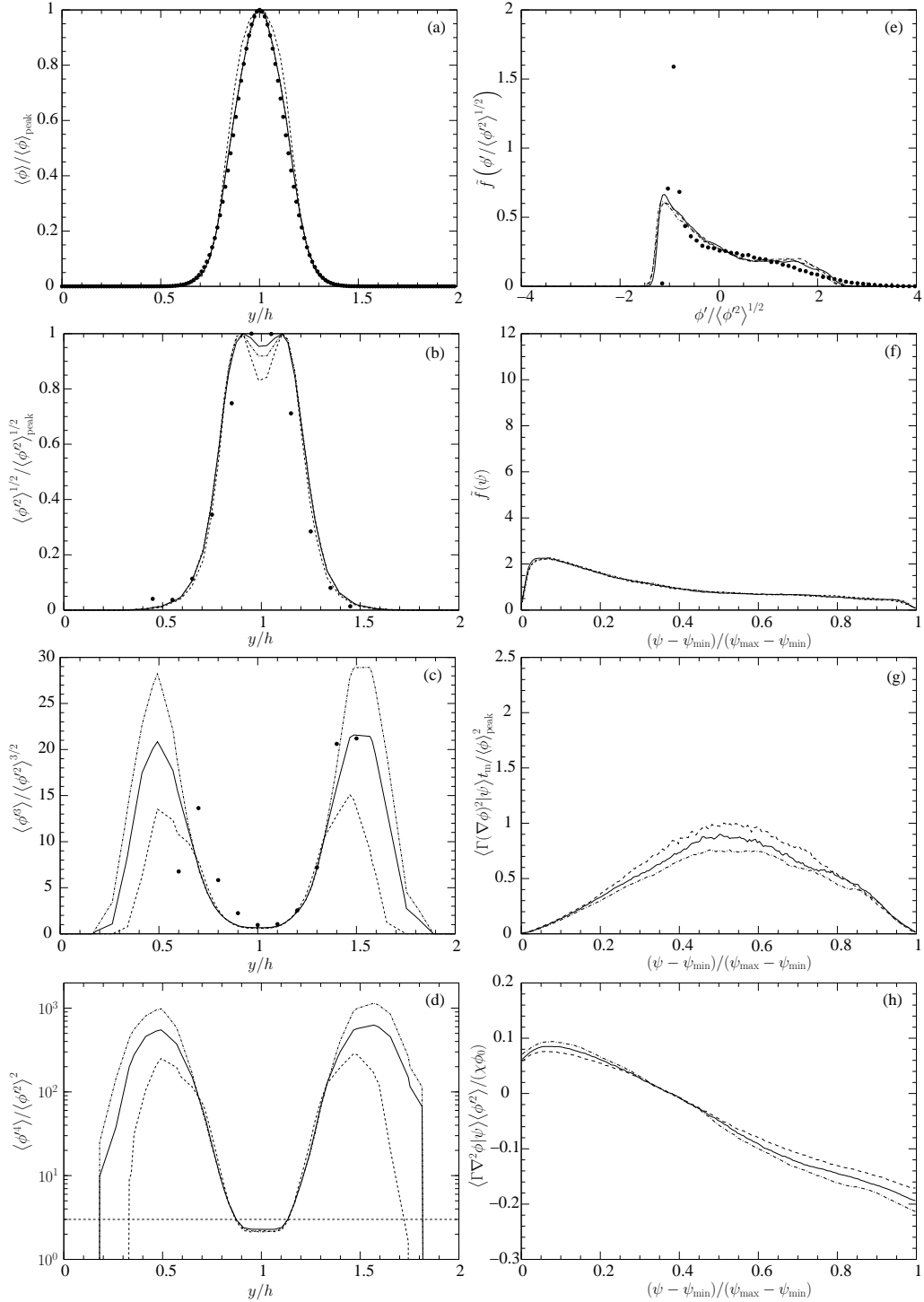

Fig. 14. Unconditional and conditional statistics of the passive scalar field affected by the number of particles with $\langle \phi | \boldsymbol{V} \rangle$ computed using the projection method of Eqs. (64)-(66) using $N_c$=5. (a)-(d) Cross-stream distribution of the first four moments at $x/h = 4.0$, (e) PDF of concentration fluctuations at $(x/h = 4.0, y/h = 1.0)$, (f) PDF of concentration at $(x/h = 4.0, y/h = 1.0)$, (g) mean scalar dissipation conditioned on the concentration at $(x/h = 4.0, y/h = 1.0)$ and (h) mean scalar diffusion conditioned on the concentration at $(x/h = 4.0, y/h = 1.0)$. Dashed line – (initial number of particles per elements) $N_{p/e}$=150, solid line – $N_{p/e}$=300 and dot-dashed line – $N_{p/e}$=500. Symbols on (a) analytical Gaussians according to Taylor (78), on (b), (c), (e) experimental data of Lavertu & Mydlarski (79). The horizontal dashed line on (d) indicates the Gaussian kurtosis value of 3.

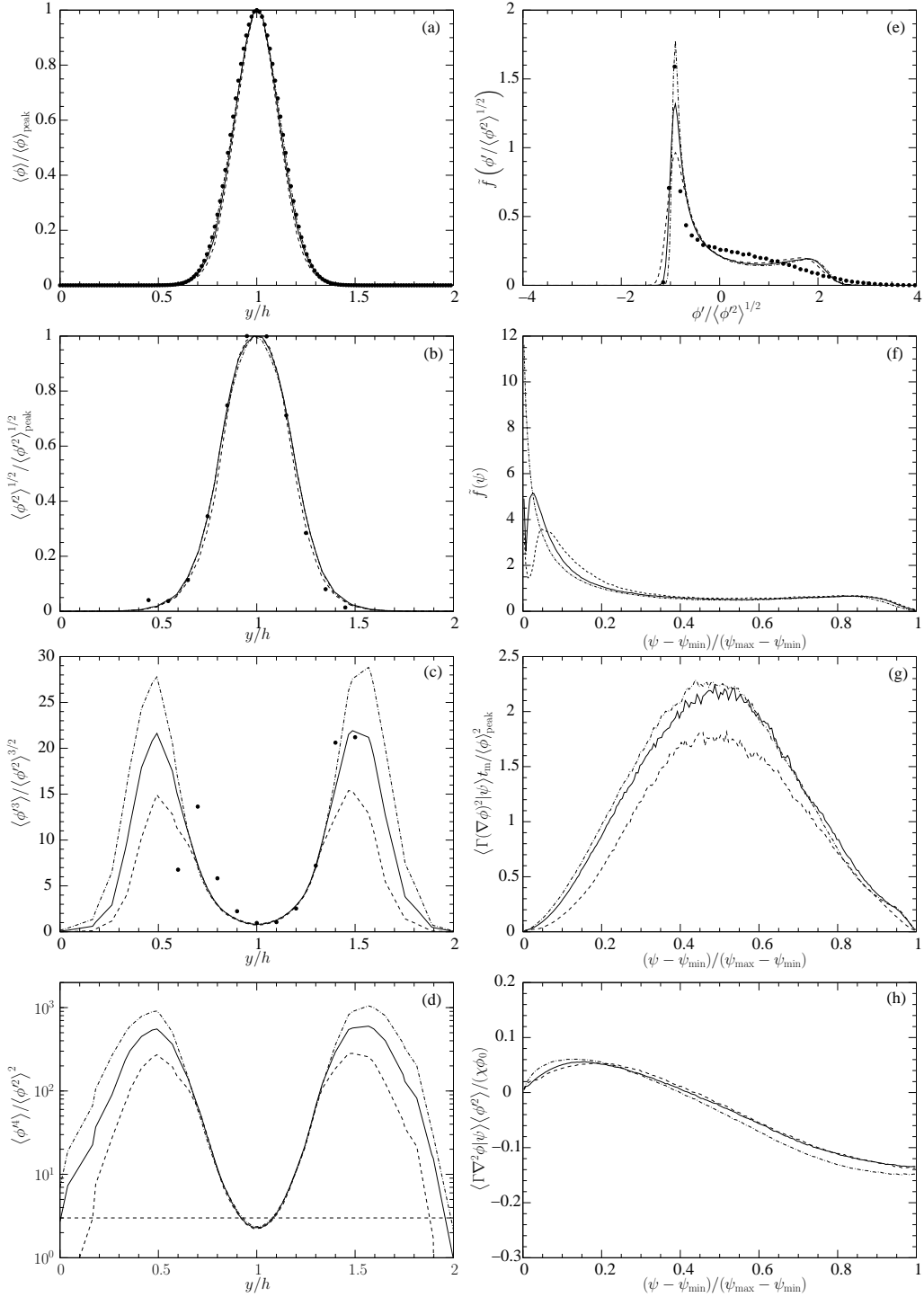

Fig. 15. Unconditional and conditional statistics of the passive scalar field affected by the number of particles with $\langle \phi | \boldsymbol{V} \rangle$ computed with the method described in Section 3.5 using a binning structure of $(5 \times 5 \times 5)$. The legend is the same as in Figure 14.

Table 5
Minimum number of particles per element required to compute different statistics.

| quantity | particles per element |
| --- | --- |
| velocity statistics, $\langle U_i \rangle, \langle u_i u_j \rangle, k, \varepsilon$ | 80–150, slightly increasing with the Reynolds number |
| first two scalar moments, $\langle \phi \rangle, \langle \phi'^2 \rangle$ | 150 |
| third, fourth and higher-order scalar moments, $\langle \phi'^3 \rangle, \langle \phi'^4 \rangle$, etc. | 500+ |
| scalar concentration PDFs, $\tilde{f}(\phi'\langle\phi'^2\rangle^{1/2}), \tilde{f}(\psi)$ | 500+ |
| mean conditional scalar dissipation, $\langle \Gamma(\nabla\phi)^2|\psi \rangle$ | 300 |
| mean conditional scalar diffusion, $\langle \Gamma\nabla^2\phi|\psi \rangle$ | 150 |

We summarize the findings for the PDF algorithm related to a passive scalar released at the centerline of a fully developed turbulent channel flow as follows:
– the prediction of one-point velocity statistics becomes more accurate with increasing Reynolds number,
– a stable numerical solution and a converged velocity field require about 80–150 particles per element depending on the Reynolds number,
– the prediction of higher-order unconditional scalar statistics and concentration fluctuation PDFs are closer to experimental observations without employing the projection technique to compute $\langle\phi|\boldsymbol{V}\rangle$,
– conditioned statistics may exhibit a large difference (up to 150%) depending on the application of the projection method, however the lack of experimental data currently prevents us to assess the true error in these quantities,
– compared to the simpler IEM model, using the IECM model only with a few conditioning intervals already makes a big difference in correcting the prediction of the scalar mean, both with and without the projection method, for an increase in the overall computational cost of about 30–40%,
– the difference in computational costs of the projection and the current general method used to compute $\langle\phi|\boldsymbol{V}\rangle$ is negligible,
– *with projection*, full convergence in the higher-order scalar statistics may require more particles than $N_{p/e}$=500, while $N_c$=3–5 was enough to reach convergence in all quantities investigated,
– *without projection*, full convergence in the higher-order scalar statistics and PDFs may require more particles than $N_{p/e}$=500, while the binning structure of $(5 \times 5 \times 5)$ was enough to reach convergent unconditional statistics, but this was still not a sufficient conditioning-resolution to achieve convergent concentration PDFs and conditional statistics.

Table 5 lists the minimum number of particles per element necessary to accurately compute the one-point statistics investigated in this study.

### 4.1.4. Computational cost

The computational cost of a simulation (the time required to reach convergence with a given accuracy) is largely determined by the resolution requirements, which in the case of a turbulent channel flow mostly amounts to adequately resolving the boundary layer. In an attempt to quantify the increase in cost of the current PDF methodology, several runs have been carried out at different Reynolds numbers between $Re_\tau = 100$ and 1080. In all cases only the statistically one-dimensional velocity field has been computed reaching a statistically stationary state, without a scalar release and micromixing. As $Re_\tau$ is increased, the boundary layer becomes thinner and a finer Eulerian grid is needed to resolve the statistics, which inevitably results in the increase of the number of particles as well. Accordingly, keeping the Courant-number approximately constant, the size of the timestep has to be decreased to achieve the same level of accuracy and stability with increasing Reynolds numbers. This tendency can be examined in Figure 16 (a), where the key factors affecting the computational cost vs. $Re_\tau$ are depicted. These are the smallest element (gridsize), the characteristic flow speed $U_c/u_\tau$, where $U_c$ is the mean velocity at the centerline, and the total number of elements $N_e$ or equivalently, the total number of particles $N_p$. All filled symbols on Figure 16 represent the given quantity normalized by the quantity at $Re_\tau = 100$. To an approximation, the number of floating-point operations, i.e. the computational cost, is proportional to the number of elements (and the number of particles) and the flow speed and inversely proportional to the gridsize (and the size of the timestep). Based on the slope of these three factors on a log-log scale, the approximate slope of the computational cost for the one-dimensional PDF simulation of channel flow can be estimated as

$$\frac{Re_\tau^{0.58} \times Re_\tau^{0.1}}{Re_\tau^{-0.88}} = Re_\tau^{1.56}, \tag{70}$$

which is in reasonable agreement with the measured $Re_\tau^{1.72}$. Using the same arguments, the cost of a three-dimensional PDF simulation can be extrapolated as

$$Re_\tau^{1.72} \times Re_\tau^{2\times 0.58} = Re_\tau^{2.88}, \tag{71}$$

which is displayed in Figure 16 (b). For comparison, the slope of the number of required elements for DNS simulations of turbulent channel flow is also displayed, based on the data reported by Abe *et al.* (76), normalized by the number of elements at $Re_\tau = 180$. This gives the slope of $Re_\tau^{2.88}$ which reasonably agrees with the prediction of Reynolds (81) for the total number of modes required as $Re_\tau^{2.7}$ ($Re_L^{2.25}$ for homogeneous turbulence (45) based on the turbulence Reynolds number). We approximate the increase in computational cost of the DNS simulations as

$$\frac{Re_\tau^{2.88} \times Re_\tau^{0.1}}{Re_\tau^{-0.88}} = Re_\tau^{3.86}. \tag{72}$$

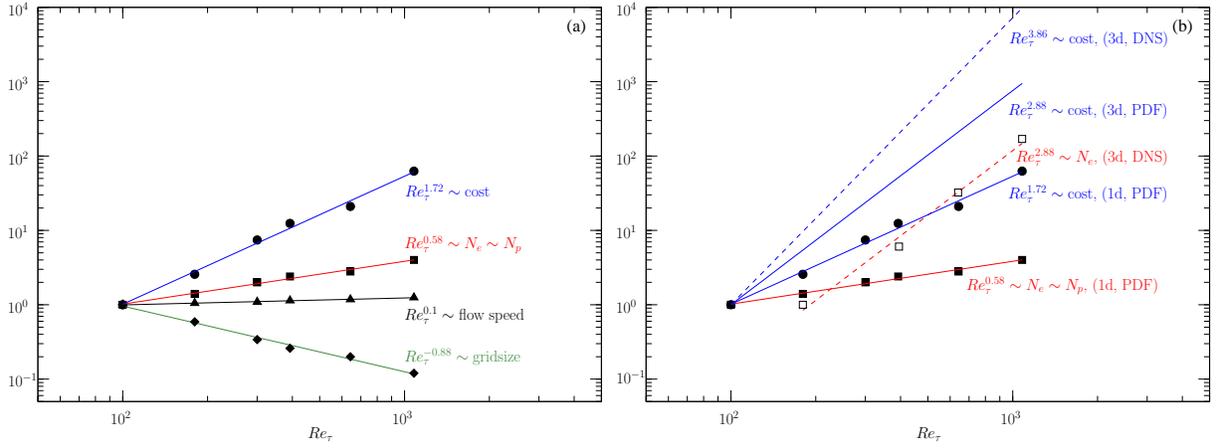

Fig. 16. Computational cost of (a) a measured one-dimensional and (b) an extrapolated three-dimensional PDF simulation. Filled symbols and solid lines – PDF calculations, hollow symbols and dashed lines – DNS of channel flow.

Now we are in a position to quantitatively compare the computational requirements of a three-dimensional PDF to DNS simulations as it is displayed in Figure 16 (b). A DNS simulation provides a great wealth of information on the turbulence for a steeply increasing cost at high Reynolds numbers. Based on Figure 16 we observe that a three-dimensional PDF simulation will probably not be as expensive for higher Reynolds numbers as DNS. As depicted in Figure 16 (b), the difference in computational cost between DNS and the three-dimensional PDF method is about a decade computing a fully resolved boundary layer at the Reynolds number $Re_\tau = 1080$. This means that at this Reynolds number DNS will produce the desired result in 10 times more computing hours than the PDF method. The figure also shows that extrapolating this result to more realistic Reynolds numbers will result in even larger differences in computational costs, DNS being increasingly more expensive than the current PDF method. As an example, resolving the boundary layer at $Re_\tau = 10^4$ will take 100 times more CPU time with DNS than with the PDF method.

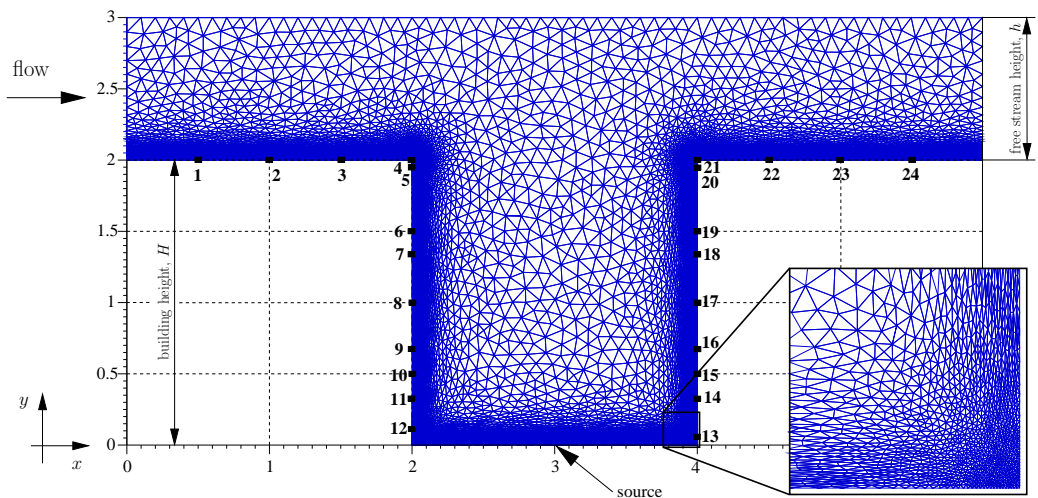

Fig. 17. Geometry and Eulerian mesh for the computation of turbulent street canyon with full resolution of the wall-boundary layers using elliptic relaxation. The grid is generated by the general purpose mesh generator Gmsh (84). The positions labeled by bold numbers indicate the sampling port locations for the passive scalar, equivalent with the combined set of measurement tapping holes of Meroney *et al.* (82) and Pavageau *et al.* (85; 86).

### 4.2. Scalar dispersion in a fully developed turbulent street canyon

The second testcase with a more complex geometry is a fully developed turbulent street canyon with a scalar released from a concentrated source at the bottom. This setup is often used to study flow patterns and pollutant dispersion in a simplified urban street canyon (82; 83). The geometry and the Eulerian grid are displayed in Figure 17. The particle copying-mirroring strategy used for the channel flow cannot be used here, so the general algorithm is applied. An additional complexity is the computation of the mean pressure in a general way, applying the pressure projection described in Section 3.2. A non-homogeneous Neumann wall-boundary condition for the pressure projection (33) has been described in Section 3.8. The flow is expected to reach a statistically steady state and is driven by a mean-pressure difference between its inflow and outflow. This condition in the free stream (above the buildings) is imposed on the mean pressure as follows.

Assuming that the inflow and outflow are aligned with $y$, as shown in Figure 17, the two-dimensional steady state cross-stream mean-momentum equation holds

Table 6
Concentration sampling locations at building walls and tops according to the experimental measurement holes of Meroney *et al.* (82) and Pavageau *et al.* (86; 85). See also Figure 17.

| #   | 1   | 2   | 3   | 4   | 5    | 6   | 7    | 8   | 9    | 10  | 11   | 12   | 13   | 14   |
|-----|-----|-----|-----|-----|------|-----|------|-----|------|-----|------|------|------|------|
| $x$ | 0.5 | 1.0 | 1.5 | 2.0 | 2.0  | 2.0 | 2.0  | 2.0 | 2.0  | 2.5 | 2.0  | 2.0  | 4.0  | 4.0  |
| $y$ | 2.0 | 2.0 | 2.0 | 2.0 | 1.93 | 1.5 | 1.33 | 1.0 | 0.67 | 0.5 | 0.33 | 0.17 | 0.17 | 0.33 |
| #   | 15  | 16   | 17  | 18   | 19  | 20   | 21  | 22  | 23  | 24  |
| $x$ | 4.0 | 4.0  | 4.0 | 4.0  | 4.0 | 4.0  | 4.0 | 4.5 | 5.0 | 5.5 |
| $y$ | 0.5 | 0.67 | 1.0 | 1.33 | 1.5 | 1.93 | 2.0 | 2.0 | 2.0 | 2.0 |

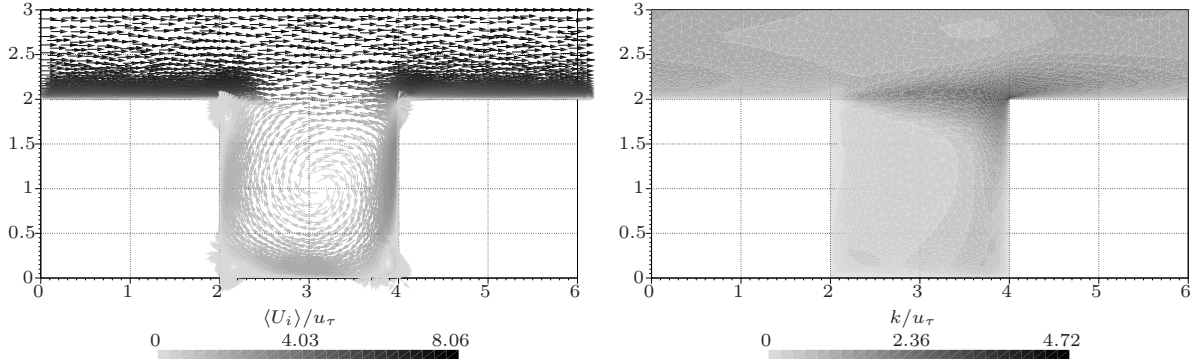

Fig. 18. Mean velocity vectors (left) and contourlines of turbulent kinetic energy (right) of a fully developed turbulent street canyon at $Re_\tau \approx 600$ based on the friction velocity and the free-stream height.

$$\frac{1}{\rho}\frac{\partial \langle P \rangle}{\partial y} = -\langle U \rangle\frac{\partial \langle V \rangle}{\partial x} - \langle V \rangle\frac{\partial \langle V \rangle}{\partial y} + \nu \left( \frac{\partial^2 \langle V \rangle}{\partial x^2} + \frac{\partial^2 \langle V \rangle}{\partial y^2} \right) - \frac{\langle uv \rangle}{\partial x} - \frac{\partial \langle v^2 \rangle}{\partial y}. \tag{73}$$

If the inflow and outflow are far enough from the canyon, the flow can be assumed to be an undisturbed turbulent channel flow. Hence we can neglect all terms on the right hand side of Eq. (73), with the exception of the last term. Thus the inflow and outflow conditions for the mean pressure can be specified according to Eq. (63). Flow-dependent non-homogeneous Dirichlet conditions have to be imposed in a way that the streamwise gradient $\partial \langle P \rangle/\partial x$ is kept at a constant level. This can be achieved by specifying the values of $\langle P \rangle$ at the inflow/outflow based on $\langle P \rangle = -\rho \langle v^2 \rangle$, which will equate their cross-stream derivatives as well. The streamwise gradient $\partial \langle P \rangle/\partial x =$ const. is applied by shifting up the values of $\langle P \rangle$ at the inflow. Consistently with Eq. (33) the above condition has to be imposed on the mean-pressure difference in time, $\delta \langle P \rangle = \langle P \rangle^{n+1} - \langle P \rangle^n$. Thus we arrive at the inflow/outflow conditions

$$\delta \langle P \rangle = \begin{cases} -\Delta P \cdot L_x - \rho \langle v^2 \rangle - \langle P \rangle^n, & \text{for inflow points}, \\ -\rho \langle v^2 \rangle - \langle P \rangle^n, & \text{for outflow points}, \end{cases} \tag{74}$$

where $\Delta P < 0$ denotes the imposed constant streamwise mean-pressure gradient over the streamwise length $L_x$ of the domain. This inflow/outflow condition drives the flow and builds up a numerical solution that converges to a statistically stationary state. No conditions are imposed on particles leaving and entering the domain other than periodicity on their streamwise positions. Wall-conditions are imposed on particles that hit wall-elements as described in Section 3.8. On the top of the domain, free-slip conditions are imposed on particles, i.e. perfect reflection on their positions and a sign reversal of their normal velocity component. To model the small-scale mixing of the passive scalar the IECM model has been applied with the $(5 \times 5 \times 5)$ binning structure without employing the projection method to compute $\langle \phi | \boldsymbol{V} \rangle$. The applied model constants for the micromixing timescale defined by Eq. (25) are the same as for the channel flow, i.e. $C_s = 0.02$ and $C_t = 0.7$.

### 4.2.1. *Results for street canyon*

The model has been run using 300 particles per element at the Reynolds number $Re_0 \approx 12000$ based on the maximum free stream velocity and the building height, $H$. Treating the free stream above the buildings as the lower part of an approximate channel flow, this corresponds to $Re_\tau \approx 600$ based on the friction velocity and the free stream height, $h = H/2$. After the flow has reached a statistically stationary state, time-averaging is used to collect velocity statistics. As an example, the mean velocity and turbulent kinetic energy fields are displayed in Figure 18.

After the flow fully develops, a passive scalar is released from a concentrated source at the center of the street level. The simulation is then run for about the same amount of (pseudo-)time as is needed to develop and time-average the velocity field. This is also a sufficient time period to collect enough samples of the released scalar, which also reaches a statistically stationary state. A wealth of statistical information is available from a PDF simulation. We report contourlines of the first four scalar moments in Figure 19.

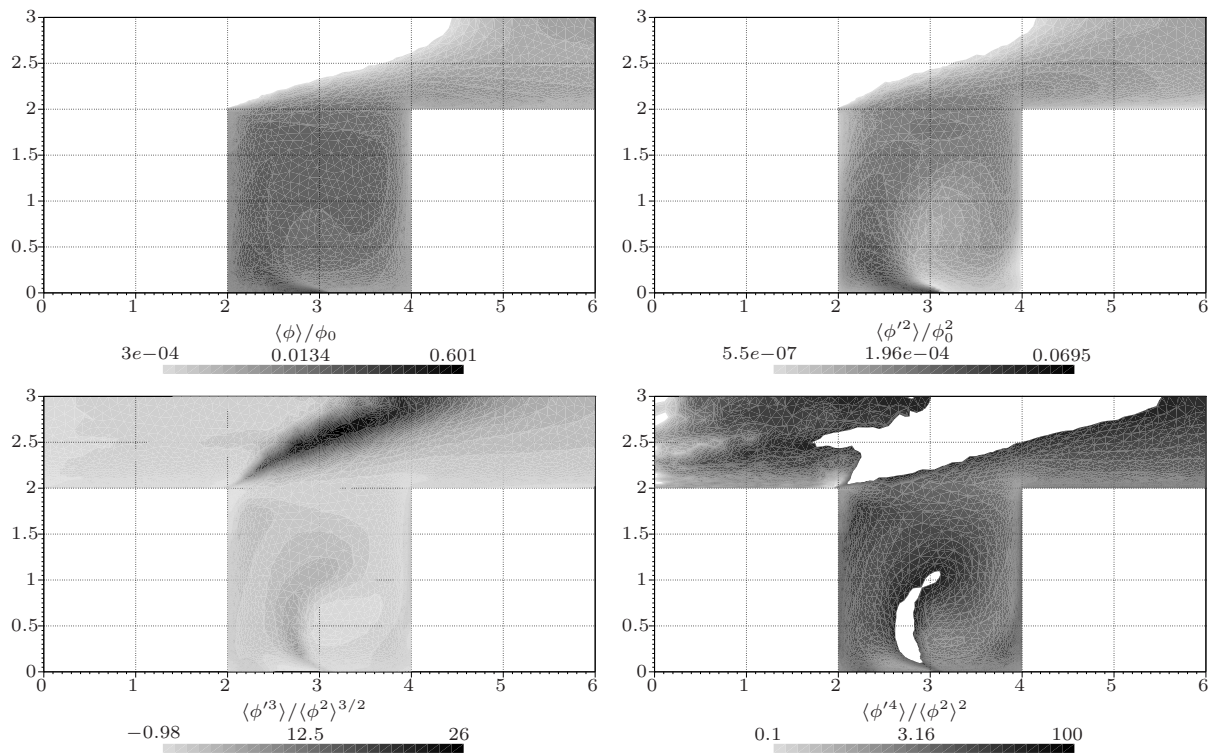

Fig. 19. The first four statistical moments of a passive scalar released at the middle of the street level of a fully developed turbulent street canyon at $Re_\tau \approx 600$. For the (a) mean, (b) variance and (d) kurtosis fields the bounds are cut to exclude extremely low values, which results in the empty (white) regions.

Several wind tunnel measurements have been carried out for this configuration, reporting concentration statistics above the buildings, on the sides of the building walls and inside the canyon (82; 85; 86). We sample the computed mean concentration field in the locations depicted and listed in Figure 17 and Table 6, respectively. Model results are plotted in Figure 20 along with a number of experiments, showing an excellent agreement.

## 5. Conclusions

This paper has presented a series of numerical methods that can be used to compute the one-point one-time joint PDF of turbulent velocity, characteristic frequency and scalar concentrations in high-Reynolds-number incompressible turbulent flows with complex geometries. Following the terminology in (20), we call the current methodology *non-hybrid* since an Eulerian CFD solver is not used in conjunction with the particle code to solve the PDF equations, i.e. the method is stand-alone. The method does belong to the familiy of particle-in-cell methods, where the Eulerian grid is used solely for: (i) estimating Eulerian statistics; (ii) tracking particles in the domain; and (iii) solving for quantities that are only represented in the Eulerian sense (i.e. mean pressure and elliptic relaxation). Compared to hybrid models, our non-hybrid method assures that none of the fields are computed redundantly, therefore the simulation is kept consistent both numerically and at the level of turbulence closure without the need to enforce consistency conditions.

Adequate wall-treatment on the higher-order statistics of the velocity field is achieved with an elliptic relaxation technique without damping or wall-functions, i.e. the boundary layers at solid (no-slip) walls are fully resolved. The examples demonstrate the applicability of the algorithm in two-dimensional flows.

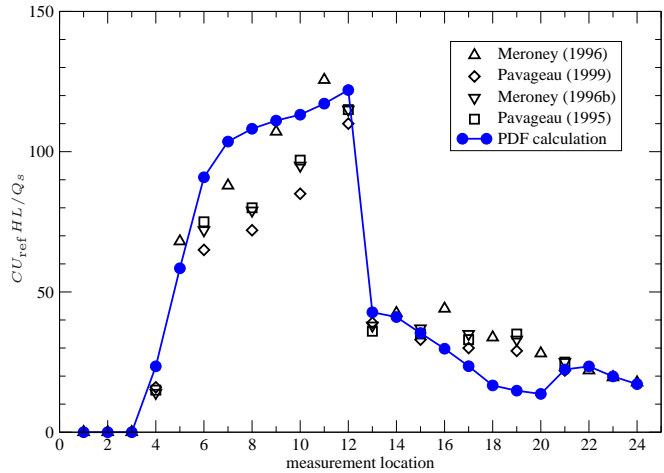

Fig. 20. Distribution of mean concentrations at the boundary of the street canyon. The experimental data are in terms of the ratio $CU_{\mathrm{ref}}HL/Q_{\mathrm{s}}$, where $C$ is the actual measured mean concentration (ppm), $U_{\mathrm{ref}}$ is the free-stream mean velocity (m/s) taken at the reference height $y_{\mathrm{ref}} \approx 11H$ and $Q_{\mathrm{s}}/L$ is the line source strength (m$^2$/s) in which $Q_{\mathrm{s}}$ denotes the scalar flow rate and $L$ is the source length. The calculation results are scaled to the concentration range of the experiments. References for experimental data: △ (82); ◇, ▽, (86); □ (85). See also Figure 17 and Table 6 for the measurement locations.

Natural future directions along these lines are the inclusion of PDF wall-functions (42) for geometrically complex flow domains and the extension to three spatial dimensions.

A significant challenge in stand-alone transported PDF methods is the accurate and stable computation of the mean pressure. This is mainly due to the following reasons: the mean velocity and Reynolds stresses have to be estimated from a noisy particle field and the pressure-Poisson equation requires their first and second derivatives, respectively, which are even noisier. We described a method to compute the mean pressure in conjunction with particle/PDF methods that only requires first derivatives of the mean velocity, which is based on a pressure-projection technique that is widely used in laminar flows.

The two Eulerian equations needed by the algorithm are both solved on unstructured Eulerian grids with the finite element method. The last couple of decades have seen great strides in automatic unstructured grid generation, grid refinement and coarsening techniques and the development of highly sophisticated grid-based data structures that minimize cache misses. Using the algorithm presented in this paper all this knowledge pertaining to unstructured meshes can be utilized in conjunction with the PDF equations and complex flow geometries. Employing finite elements together with particle/PDF methods also has the advantage of greatly simplifying boundary conditions for particles – no ghost elements are required as in finite volume methods. Furthermore, finite element approximation functions are not only used for particle tracking but also provide an elegant way of estimating derivatives of statistics from particle fields.

We also described a general algorithm that can be used to calculate the velocity-conditioned scalar mean for the IECM micromixing model. The procedure homogenizes the statistical error over the sample space for arbitrary velocity PDFs by dynamically adjusting the number of bins and their distribution. A particle-redistribution algorithm has also been described that provides stability by ensuring that no Eulerian elements remain without particles at any time during the simulation.

We also proposed a general form for the micromixing timescale that can be used in a flow-, and geometry-independent manner for modeling the effect of small-scale mixing on a transported passive scalar released from a concentrated source. Although the computed concentration results compare well with analytical and experimental data for the two testcases, no final conclusions can be drawn regarding the most suitable mathematical expression and modeling constants.

Regarding computational costs, Pope (45) places PDF methods somewhere between Reynolds stress closures and large eddy simulation. The solver has been parallelized with the OpenMP standard, which easily allows the exploitation of multicore workstations mainly used for production codes. Our performance study has shown a good parallel speedup up to 32 CPUs tested on shared memory machines using single-, dual-, and quad-core CPUs. We also ported the code to Intel's Cluster OpenMP technology, which allows an OpenMP program to run on a beowulf-type cluster of networked workstations requiring a minor programing effort compared to an MPI-based implementation. However, we found that the algorithm with its current design is not suitable for Cluster OpenMP.

The fields calculated for testcases show a good agreement when compared to DNS and experimental data where available. In the future, further testing with cases of different complexity will be carried out.

## 6. Appendix A

In Section 3.5 a numerical strategy to estimate the velocity-conditioned scalar mean $\langle\phi|\boldsymbol{V}\rangle$ required in Eq. (24) is detailed. An algorithm that accomplishes the conditioning step after the particles have been sorted in element e into subgroups may be written as follows. Let $\texttt{CNBI}(=N_c)$, $\texttt{NELEM}(=N_e)$, $\texttt{NPAR}(=N_p)$ and $\texttt{MAXNPEL}(=N_{p/e}^{\max})$ denote the number of conditioning bins, the total number of elements of the Eulerian grid, the total number of particles and the maximum number of particles per elements, respectively. Furthermore, let the arrays np[CNBI], vcce[NELEM*CNBI], npel[NELEM], parid[MAXNPEL] and parc[NPAR] represent the number of particles in bins, the velocity-conditioned scalar concentration in bins of each element, the actual number of particles in each element, the indices of the particles residing in element e and the particle concentrations, respectively. (Note the use of C-style indexing, i.e. the array indices start from 0.)

Sort parid[0:MAXNPEL-1] according to the sorting & dividing procedure described in Section 3.5

Initialize np[0:CNBI-1] = vcce[0:CNBI-1] = n = 0;

```
for all particles in element e
  i = CNBI*n/npel[e];                                 // compute bin index
  np[i] = np[i] + 1;                                  // increase number of particles in bin i
  vcce[e*CNBI+i] = vcce[e*CBI+i] + parc[parid[n]];   // add particle concentration to bin i
  cp[parid[n]] = i;                                   // store conditioning pointer for particle
  n = n + 1;                                          // increase number of particles considered
end

for all bin i
  vcce[e*CNBI+i] = vcce[e*CNBI+i]/np[i];             // finish computing conditional mean in bin i
end
```

After this algorithm, the array cp[NPAR] will contain conditioning pointers for each particle relative to their host element, so that the velocity-conditioned scalar mean for particle p in element e can be obtained as vcce[e*CNBI+cp[p]].

## 7. Appendix B

In Section 3.7 the need for a particle-redistribution algorithm is emphasized. What follows is the algorithm that we employ in order to keep the number of particles per element above a certain treshold.

```
do {

    find the elements (mine, maxe) containing the
        smallest and largest number of particles (minnpel, maxnpel);

    if { (minnpel < MINNPEL) and (minnpel ≠ maxnpel) }
      move a particle from element maxe to mine;

} while { (minnpel < MINNPEL) and (minnpel ≠ maxnpel) };
```

The loop stops if the required minimum number of particles per element $N_{p/e}^{\min}$ is reached or the element-distribution of particles becomes homogeneous on the domain. Any particle may be moved from element maxe to mine as long as the local statistics are not altered. In principle, this can be achieved if the properties $(\mathcal{U}_i, \omega, \psi)$ of the newly arriving particle in element mine are sampled from the local joint PDF. A quick way to do this is to initialize the particle properties by copying another (randomly chosen) one already residing in element `mine`. Since the joint PDF is represented by a finite number of particles, taking out a particle from element `maxe` and putting it into `mine` will alter the local statistics in both elements, even if the new properties are copied from a neighbor. Since `maxe` contained the largest number of particles on the whole domain, we are less concerned about the effect of a single leaving particle since the local PDF is well represented there. However, the effect of the newly introduced particle in element `mine` where the joint PDF was already poorly represented is of higher importance. Thus in Section 3.7 we investigate the error introduced by the particle redistribution using a simplified governing equation.